\documentclass[twocolumn,preprintnumbers,amsmath,amssymb]{revtex4}

\usepackage{graphicx}
\usepackage{dcolumn}
\usepackage{bm}

\usepackage{xcolor}
\usepackage{placeins}
\usepackage{ulem}

\newcommand{\dima}[1]{\textcolor{red}{DY: #1}}

\begin{document}

\title{Optical second and third harmonic generation on excitons\\ in ZnSe/BeTe quantum wells}

\author{Johannes Mund$^1$, Andreas Farenbruch$^1$, Dmitri R. Yakovlev$^{1,2}$, Andrei A. Maksimov$^3$, Andreas Waag$^4$, and Manfred Bayer$^{1,2}$}
\affiliation{$^1$Experimentelle Physik 2, Technische Universit\"at Dortmund, D-44221 Dortmund, Germany}
\affiliation{$^2$Ioffe Institute, Russian Academy of Sciences, 194021 St. Petersburg, Russia}
\affiliation{$^3$Institute of Solid State Physics, Russian Academy of Sciences, 142432 Chernogolovka, Russia}
\affiliation{$^4$Institute of Semiconductor Technology, Braunschweig Technical University, 38106 Braunschweig, Germany}


\begin{abstract}
Optical harmonic generation on excitons is found in ZnSe/BeTe quantum wells with type-II band alignment. Two experimental approaches with spectrally-broad femtosecond laser pulses and spectrally-narrow picosecond pulses are used for spectroscopic studies by means of second and third harmonic generation (SHG and THG). The SHG signal is symmetry-forbidden for light propagation along the structure's growth axis, which is the [001] crystal axis, but can be induced by an external magnetic field. The THG signal is detected at zero magnetic field and its intensity is field independent. A group theory analysis of SHG and THG rotational anisotropy diagrams allows us to identify the involved excitation mechanisms.
\end{abstract}


\maketitle

\section{Introduction}
\label{sec.introduction}

Nonlinear optical spectroscopy is a powerful technique to study electronic states in solids~\cite{Shen_book,Boyd_book}. Among the variety of the used approaches multi-photon processes, like optical second harmonic generation (SHG), are of particular interest as they give access to the symmetry of the electronic states involved in the optical transitions~\cite{Fiebig05,Pisarev10}. Using magnetic- or electric-field-induced SHG one can address optical transitions that are symmetry-forbidden for linear optical techniques. The reason is that SHG is a coherent process involving optical transitions that have to be allowed both for two-photon excitation and one-photon emission. Optical third harmonic generation (THG), which involves three ingoing photons and one outgoing photon, offers new degrees of freedom to study electronic states. For example, in electric-dipole approximation SHG is symmetry-forbidden in crystals with a center of inversion, whereas THG is possible in both centrosymmetric and non-centrosymmetric crystals.          

Systematic spectroscopic studies of excitons in semiconductors by optical harmonic generation started in 2004, see Refs.~\cite{Pisarev10,Yakovlev18} and references therein. In these studies, the fundamental laser is tuned to a frequency $\omega$ in the transparency range such that the double and triple frequencies, $2\omega$ and $3\omega$, of the optical harmonics are in resonance with exciton states. Resonant enhancement of the SHG and THG on the exciton states, also in magnetic field, was found. Several groups of bulk semiconductors were studied, such as diamagnetic GaAs, CdTe, ZnSe, ZnO, Cu$_2$O, and diluted magnetic (Cd,Mn)Te. It was shown that the symmetry reduction of exciton states in external electric and/or magnetic fields can activate them in SHG and substantially enhance their contribution to THG. New mechanisms of optical harmonic generation, specific for excitons and exciton-polaritons, were disclosed and a microscopic consideration of these mechanisms was carried out~\cite{Lafrentz13,Brunne15,Warkentin18,Mund18}. Recently, SHG spectroscopy was extended to two-dimensional semiconductors MoS$_2$ and  WSe$_2$~\cite{Wang15,Trolle15,Glazov17}.

Despite these advances, so far SHG/THG spectroscopy has not been used for investigation of excitons confined in semiconductor heterostructures. We published only preliminary data for ZnSe/BeTe multiple quantum wells (MQW) demonstrating that it is feasible in principle~\cite{Yakovlev18}. Here, we present a systematic spectroscopic study of SHG and THG on the direct excitons in ZnSe/BeTe MQW with a type-II band alignment. We choose a ZnSe-based MQW due to the pronounced exciton resonances in optical spectra and the relatively large exciton binding energy exceeding 20~meV, and therefore the large oscillator strength~\cite{Platonov98}. Due to the 20~nm thick ZnSe quantum well layers, exceeding the exciton Bohr radius of 4.5~nm, the direct exciton in the ZnSe/BeTe MQW forms a stable complex at low temperatures with confinement in the ZnSe layers. In detail, the electrons are confined in the ZnSe layers by the heterostructure potential, while the holes, which have their potential minimum in the BeTe layers, take on metastable states in the ZnSe layers due to the Coulomb attraction by the electrons~\cite{Platonov98,Maksimov99}. 

An interesting property of type-II quantum wells is the spatial separation of electrons and holes, which, in case of ZnSe/BeTe structures, can strongly modify the confinement potentials, resulting in a change of the spectral properties of spatially direct and indirect emissions, see Fig.~\ref{pic.sample}, and of their recombination dynamics~\cite{Zaitsev97,Maksimov08}. ZnSe and BeTe have no common ions, thus enhancing the anisotropy effects caused by the layer interfaces. This anisotropy is particularly pronounced for the indirect optical transitions~\cite{Platonov99,Yakovlev00}, but also can have some contribution to the direct transitions, especially when the two opposite interfaces in the QW are formed by Be-Se and Zn-Te bonds, respectively~\cite{Zaitsev02}.    

The paper is organized as follows. In Sec.~\ref{sec.experiment} details of the sample and experimental techniques are given. Experimental data on SHG and THG measured by spectrally broad fs laser pulses and scanning of ps laser pulses, as well as the observed signal from the GaAs substrate are described in Sec.~\ref{sec.results}. Section~\ref{sec.discussion} contains discussion of the data and model considerations of the rotational anisotropies of the SHG and THG signals.  

\section{Experimental details}
\label{sec.experiment}

\begin{figure}[h]
\begin{center}
	\includegraphics[width=0.35\textwidth]{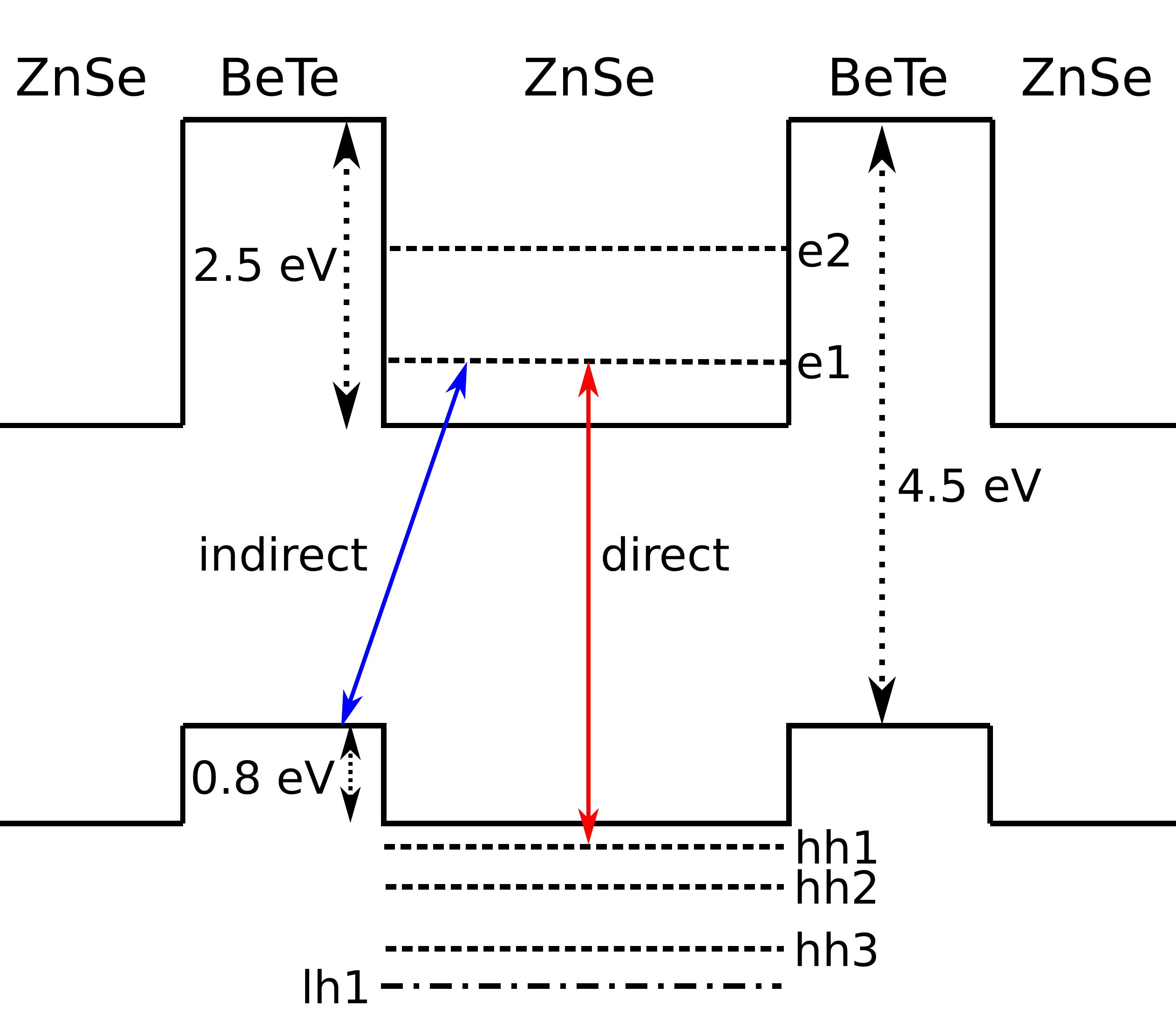}
	\caption{Band diagram of the ZnSe/BeTe MQW structure with type-II band alignment. The electron states confined in the ZnSe layer (e1, e2) and the above-barrier states of the heavy-hole (hh1, hh2, hh3) and light-hole (lh1)  are shown by the dashed and dash-dotted lines.}
	\label{pic.sample}
\end{center}
\end{figure}

The studied ZnSe/BeTe multiple quantum well structure (cb1750) was grown by molecular-beam epitaxy on a 0.5-mm-thick GaAs substrate with [001] orientation. It consists of ten periods, each consisting of a 20-nm-thick ZnSe layer and a 10-nm-thick BeTe layer. The growth conditions were adjusted to form Zn-Te chemical bonds at all interfaces~\cite{Platonov99,Yakovlev00,Zaitsev02}. In this case, the Zn-Te and Te-Zn chemical bonds at opposite interfaces are oriented perpendicularly to each other. At low temperatures, ZnSe has a band gap energy of 2.82~eV at the $\Gamma$-point of the Brillouin zone, and the band gap energy of BeTe is about 4.5~eV~\cite{Nagelstrasser98}. The ZnSe/BeTe material system has a type-II band alignment with a very large confinement potential for the electrons of 2.5~eV, so that the electrons are very well localized in the ZnSe layers. The confinement potential of the holes is 0.8~eV, and the minimum of their energy is located in the BeTe layers. The band diagram is shown schematically in Fig.~\ref{pic.sample}. The direct and indirect optical transitions in real space have energies of about 2.8~eV and 2.0~eV, respectively. Both are seen in the photoluminescence at low temperatures~\cite{Platonov98,Maksimov99,Maksimov08}. In this study, we focus on the direct optical transition only, whose exciton resonances contribute to the SHG and THG signals. Since the ZnSe layers are relatively thick as compared to the exciton radii the exciton binding energy is close to the ZnSe bulk value of 20~meV.  

\begin{figure}[h]
\begin{center}
	\includegraphics[width=0.49\textwidth]{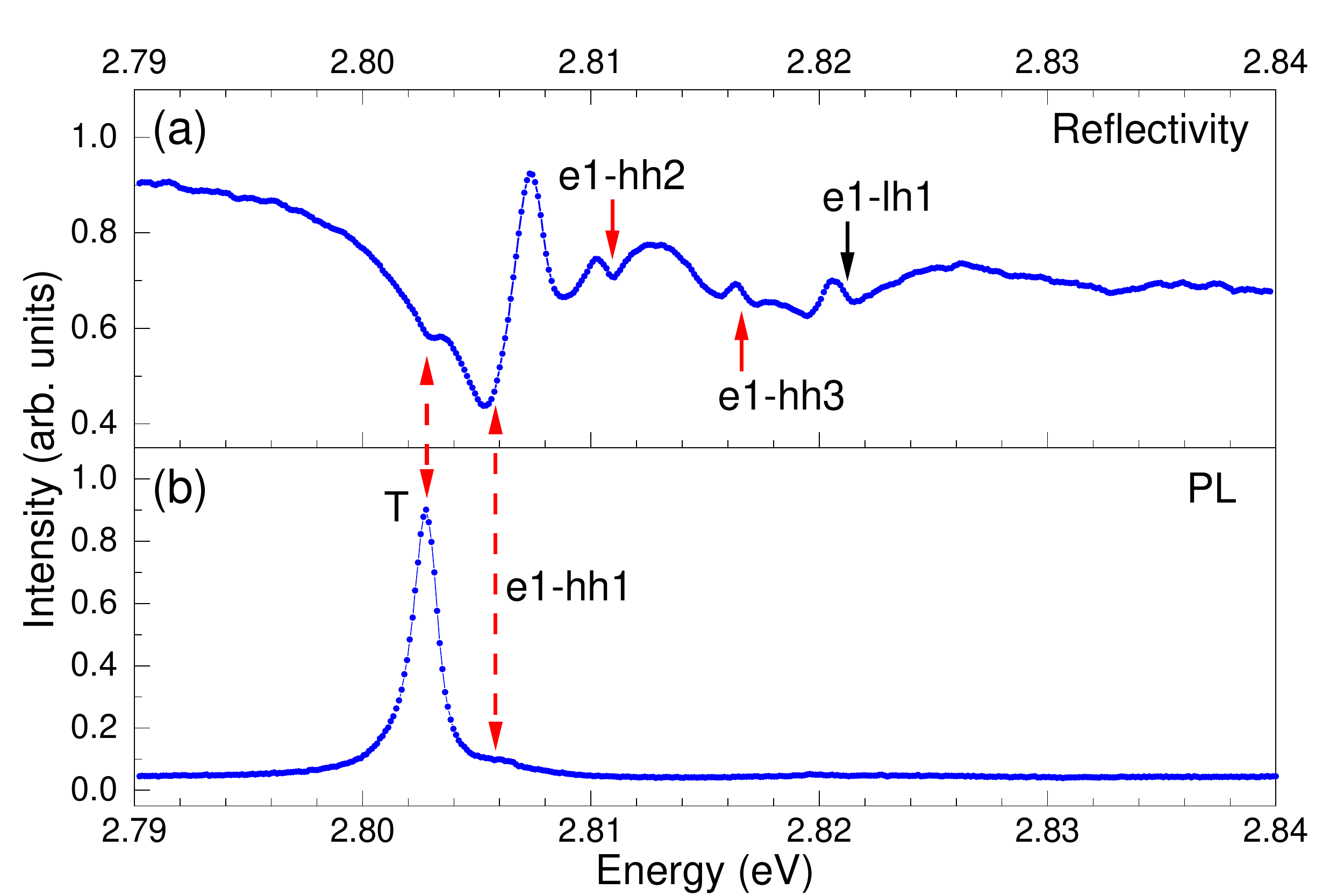}
	\caption{Linear optical spectra of the ZnSe/BeTe MQW structure measured at $T=5$~K. (a) Reflectivity spectrum. Red and black arrows indicate the resonance energies of the 1S transitions between the lowest electron level e1 and different heavy-hole (hhi) and light-hole (lh) levels, respectively. (b) Photoluminescence spectrum excited with 3.06~eV laser photon energy. }
	\label{pic.Refl+PL}
\end{center}
\end{figure}

We measured linear optical spectra (reflectivity and photoluminescence) of the studied structure in order to have a reference for comparison with the SHG/THG spectra. These spectra recorded at the temperature of $T=5$~K are shown in Fig.~\ref{pic.Refl+PL}. The reflectivity was measured in a backscattering geometry with the use of a halogen lamp. In Fig.~\ref{pic.Refl+PL}(a), the strongest resonance at 2.8059~eV, labelled as e1-hh1, corresponds to the exciton formed from the confined electron ground state (e1) and the ground heavy-hole state (hh1). We identify the two further resonances on the high energy side as e1-hh2 and e1-hh3, involving transitions from the fist and second excited heavy-hole state. The resonance at 2.821~eV is assigned to the e1-lh1 transition involving the ground light-hole state. The energy splitting between the hh1 and lh1 states of about 15~meV is provided by strain, induced by a small lattice mismatch between the ZnSe and BeTe layers. 

The photoluminescence (PL) was excited by a continuous-wave laser with a photon energy of 3.06~eV, providing an excitation density of 7~W/cm$^2$. In the PL spectrum, shown in Fig.~\ref{pic.Refl+PL}(b), one sees the emission of the e1-hh1 neutral exciton only as a shoulder at 2.8059~eV, which is located on the high energy flank of the strongest PL line at 2.8028~eV. This emission feature is the emission of the negatively charged excitons [trions (T)]~\cite{Debus10}. It can be seen also as a weak resonance in the reflectivity spectrum. The additional electrons needed for the trion formation are provided in the ZnSe layers by photogeneration due to scattering of holes into the BeTe layers. Due to their spatial and energetic separation by the large band offsets, these electrons and holes have very long recombination times because of the underlying indirect optical transition.  

\begin{figure}[h]
\begin{center}
	\includegraphics[width=0.4\textwidth]{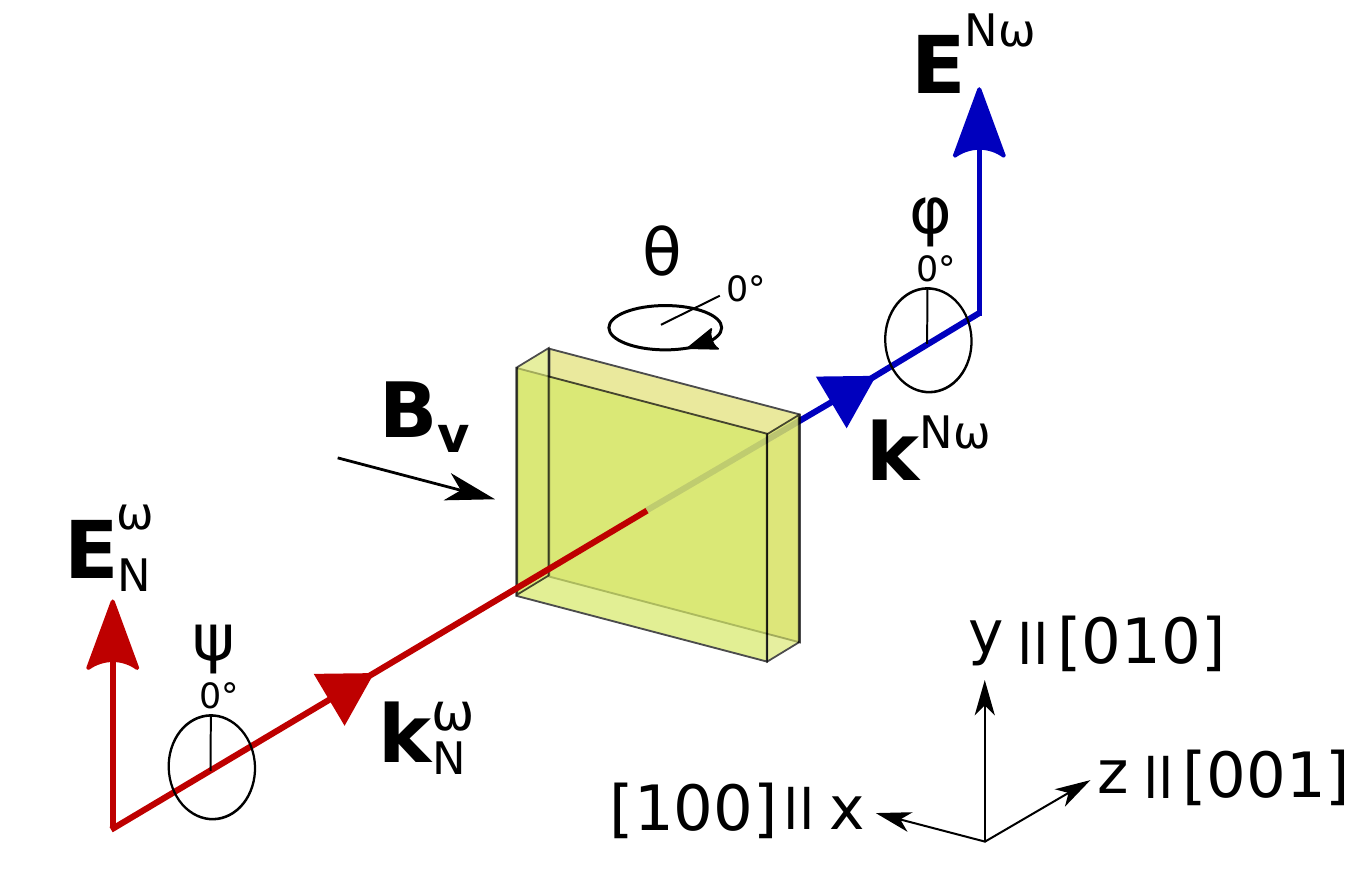}
	\caption{Sample orientation relative to the optical axis and linear polarization angles in the optical harmonic generation experiments. The incoming light with frequency $\omega$ and electric field $\textbf{E}^\omega_N$ has the wave-vector $\textbf{k}^\omega_N$ and the linear polarization angle $\psi$. The generated harmonic light with frequency $N\omega$ ($N=2,3$) and electric field  $\textbf{E}^{N\omega}$ has the wave vector $\textbf{k}^{N\omega}$ and is detected at the polarization angle $\varphi$. The magnetic field $\textbf{B}_\mathrm{V}$ is applied along the crystal [100]-direction, perpendicular to the funsamental light direction $\textbf{k}^\omega$ (Voigt geometry).}
	\label{pic.setup}
\end{center}
\end{figure}

For the SHG/THG experiments, we use a recently developed technique for exciton spectroscopy based on optical harmonic generation through laser excitation with fs laser pulses followed by detection  with high spectral resolution. The details of the technique can be found in Refs.~\cite{Mund18,Farenbruch20}. In Fig.~\ref{pic.setup}, the sample orientation relative to the optical axis as well as the linear polarization angles of the fundamental and harmonic light are specified.

The pump laser in our setup emits pulses of 150~fs duration at a repetition rate of 30~kHz, pumping two optical parametric amplifiers (OPA) of which one emits pulses of 3.3~ps duration with a full width at half maximum (FWHM) of about 0.6~meV. The other OPA emits pulses of 200~fs duration with a FWHM of about 10~meV. The OPA central photon energy can be tuned in the range of relevance for the optical harmonic generation around $E_g/N$, where $E_g=2.82$~eV is the ZnSe band gap energy at cryogenic temperatures and $N=2$ and 3 for SHG and THG, respectively. The energy per pulse is set to $0.1-1.0~\mu$J for both OPAs depending on the harmonic order to be measured.

The laser beam hits the ZnSe/BeTe sample surface under normal incidence parallel to the [001] crystal direction. It is focused into a spot with a diameter of about 100~$\mu$m. The signals are detected by the combination of a spectrometer and a silicon charge-coupled device (CCD) camera (matrix with $1340 \times 400$ pixels, pixel size 20~$\mu$m), cooled by liquid nitrogen. The spectrometer is a 0.5-meter focal length monochromator (Acton, Roper Scientific) with a 1800 grooves/mm grating. The overall spectral resolution of the detection system at photon energies around 2.8~eV is about 100~$\mu$eV.

Using a Glan-Thompson polarizer and a half-wave plate, the linear polarization plane of the ingoing and outgoing light can be rotated continuously and independently. One can thus detect the signals for any chosen relative polarization of $\textbf{E}^\omega$ or $\textbf{E}^{N\omega}$ and, therefore, measure the rotational anisotropy diagrams of the optical harmonics. Here, we measure these anisotropies for either parallel ($\textbf{E}^\omega\parallel\textbf{E}^{N\omega}$) or crossed ($\textbf{E}^\omega\perp\textbf{E}^{N\omega}$) linear polarizations of the laser and the harmonic light. The analyzer is set to the optimal polarization, providing highest throughput through the monochromator. A long pass filter placed before the sample prevents SHG by optical elements from entering the monochromator and a short pass filter placed after the sample is used for cutting off the infrared pump light.

The optical harmonic generation spectra are measured by exciting the sample with laser pulses emitted by either the fs-OPA or the ps-OPA. With the spectrally-broad fs-pulses the whole spectral range around the 1S exciton in ZnSe layers with the resonance energy  ${\cal E}_\textrm{1S}$ can be excited. For that the central photon energy of the OPA output was set to ${\cal E}_\textrm{1S}/N$ with the necessary $N$ for SHG or THG. When using the spectrally-narrow ps-pulses the photon energy has to be tuned across the energy range of the resonances of interest, divided by $N$. Therefore, it is necessary to detect the harmonic signal for each photon energy separately.

For measurements in magnetic field, we use a superconducting split-coil magnet (Oxford Instruments) allowing field strengths up to 10~T. The field is directed along the [100] crystal direction, perpendicular to the light $k$-vector (Voigt geometry). For optical measurements, the sample is kept in a bath cryostat at a temperature of $T = 5$~K in contact with cold helium gas.

\section{Experimental Results}
\label{sec.results}

The outline for presenting the experimental data is the following. We start with THG spectra at zero magnetic field and then turn to THG and SHG in external magnetic field. Further insight into the exciton resonances is obtained by measuring the THG and SHG rotational diagrams. Tilting the sample ($\textbf{k}^\omega\nparallel[001]$) allows us to measure crystallographic SHG signal from the GaAs substrate and detect its absorption by QW excitons.

\subsection{Third harmonic generation}
\label{sec.THG}

\begin{figure}[h]
\begin{center}
	\includegraphics[width=0.49\textwidth]{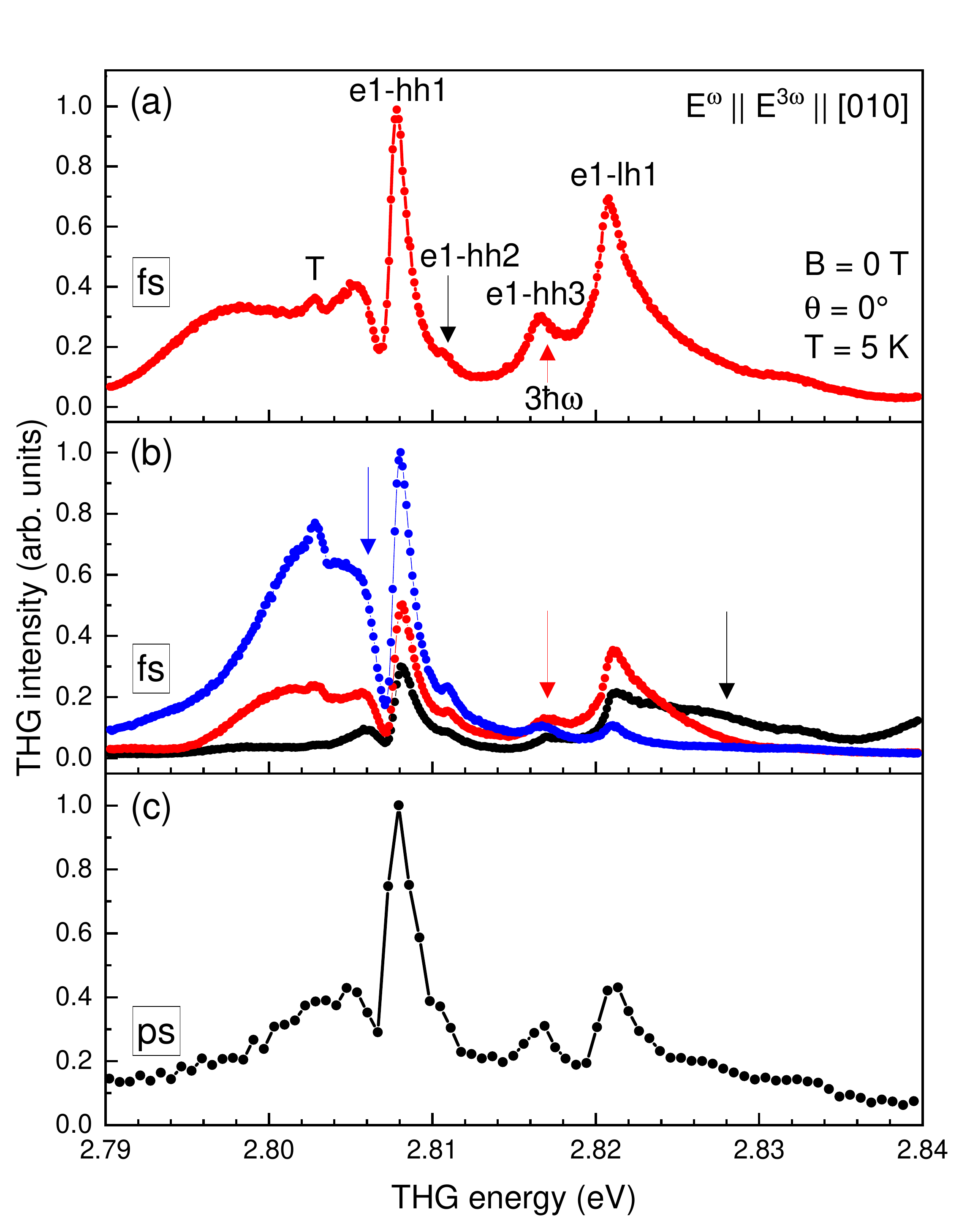}
	\caption{THG spectra of ZnSe/BeTe MQW measured at $B=0$~T and $T=5$~K with normal light incidence ($\theta=0^\circ$, $\textbf{k}^\omega\parallel [001]$). Polarizations are $\textbf{E}^\omega\parallel\textbf{E}^{3\omega}\parallel[010]$. (a) Spectrum measured with fs pulses with maximum energy at $3\hbar\omega=2.817$~eV. (b) Spectra with fs-pulses centered at $3\hbar\omega=2.806$~eV (blue), 2.817~eV (red) and 2.828~eV (black). (c) Spectrum obtained by scanning the ps-laser excitation energy. }
	\label{pic.[001]_THG_all_ps+fs}
\end{center}
\end{figure}

In Figure~\ref{pic.[001]_THG_all_ps+fs}, THG spectra measured with spectrally broad fs-pulses and spectrally narrow ps-pulses are compared. In Figure~\ref{pic.[001]_THG_all_ps+fs}(a), the fs-pulse is centered at $3\hbar\omega=2.817$~eV and extends over the whole presented spectral range. The laser pulse has a smooth shape, while the THG spectrum contains sharp spectral features. The energies of these features correspond well to the energies of the exciton states observed in reflectivity, compare with Fig.~\ref{pic.Refl+PL}(a). The e1-hh1 1S resonance has the highest amplitude and a FWHM of 1.1~meV. The relative intensities of the exciton resonances in the THG spectra obviously depend on the central laser photon energy, as shown in Fig.~\ref{pic.[001]_THG_all_ps+fs}(b). Here, three spectra recorded for different central energies marked by the arrows are given. The relative intensities between the resonances become redistributed, but the resonance energies remain the same. This additionally confirms that the spectral features are not caused by possible spectral modulation of the fs-laser pulses. 

\FloatBarrier

We note that by the fs-pulses the whole spectral range, shown in Fig.~\ref{pic.[001]_THG_all_ps+fs}, is covered without scanning the central laser energy. Therefore, only five minutes accumulation time is needed to measure the spectrum with a good signal-to-noise ratio. Thus, THG spectroscopy with fs-pulses is a valuable tool to quickly identify resonances and their spectral widths. To quantitatively measure the relative peak intensities, however, requires normalization to the fs-laser spectral intensity, accounting for the optical nonlinearity. The THG spectrum measured by scanning the ps-laser is shown in Fig.~\ref{pic.[001]_THG_all_ps+fs}(c). The ps-laser intensity is about constant in this spectral range and normalization is not needed to assess the relative intensities of the exciton resonances. One can observe in this spectrum all resonances detected also under fs-pulses, which are spectrally broadened here due to the limited resolution of the ps-pulse FWHM of 0.6~meV. The acquisition time of the whole spectrum is 15 minutes with accumulation time at each data point of 10 seconds, the signal-to-noise ratio is not as good as for fs-pulses. 

\begin{figure}[h]
\begin{center}
	\includegraphics[width=0.5\textwidth]{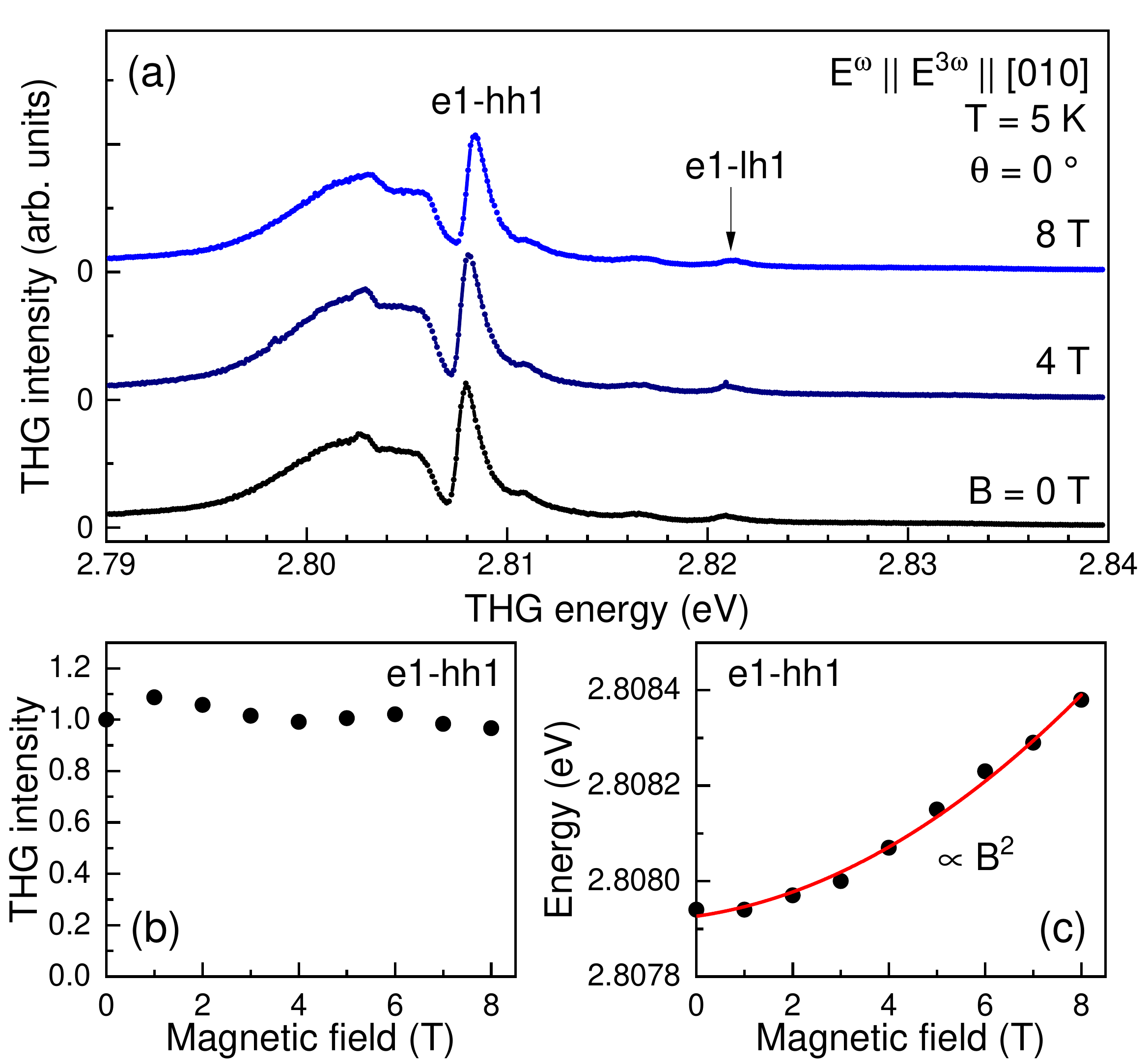}
	\caption{(a) THG spectra for different magnetic fields. Sample is oriented for normal light incidence $\theta=0^\circ$ and polarizations are $\textbf{E}^\omega\parallel\textbf{E}^{3\omega}\parallel[010]$. The magnetic field dependencies of the THG intensity and peak energy of the 1S exciton are shown in panels (b) and (c), respectively.}
	\label{pic.[001]_THG_all_0-8Tv_fs}
\end{center}
\end{figure}

As one can see in Fig.~\ref{pic.[001]_THG_all_0-8Tv_fs}(a), application of an external magnetic field up to 8~T has only small influence on the THG spectra. The THG intensity remains about constant, as is shown for the e1-hh1 resonance in Fig.~\ref{pic.[001]_THG_all_0-8Tv_fs}(b). A similar behavior was reported previously for bulk ZnSe~\cite{Warkentin18}. It was explained by the relatively small exciton diameter in ZnSe and, therefore, magnetic fields up to 8~T do not shrink the wave function considerably so that the exciton oscillator strength remains about constant. This is in correspondence with the weak diamagnetic shift of the resonance energy of the 1S exciton from the e1-hh1 transition in the THG spectra[Fig.~\ref{pic.[001]_THG_all_0-8Tv_fs}(c)]. It has characteristic $B^2$ dependence and reaches 0.4~meV at $B=8$~T. 

\FloatBarrier

\begin{figure}[h]
	\begin{center}
		\includegraphics[width=0.5\textwidth]{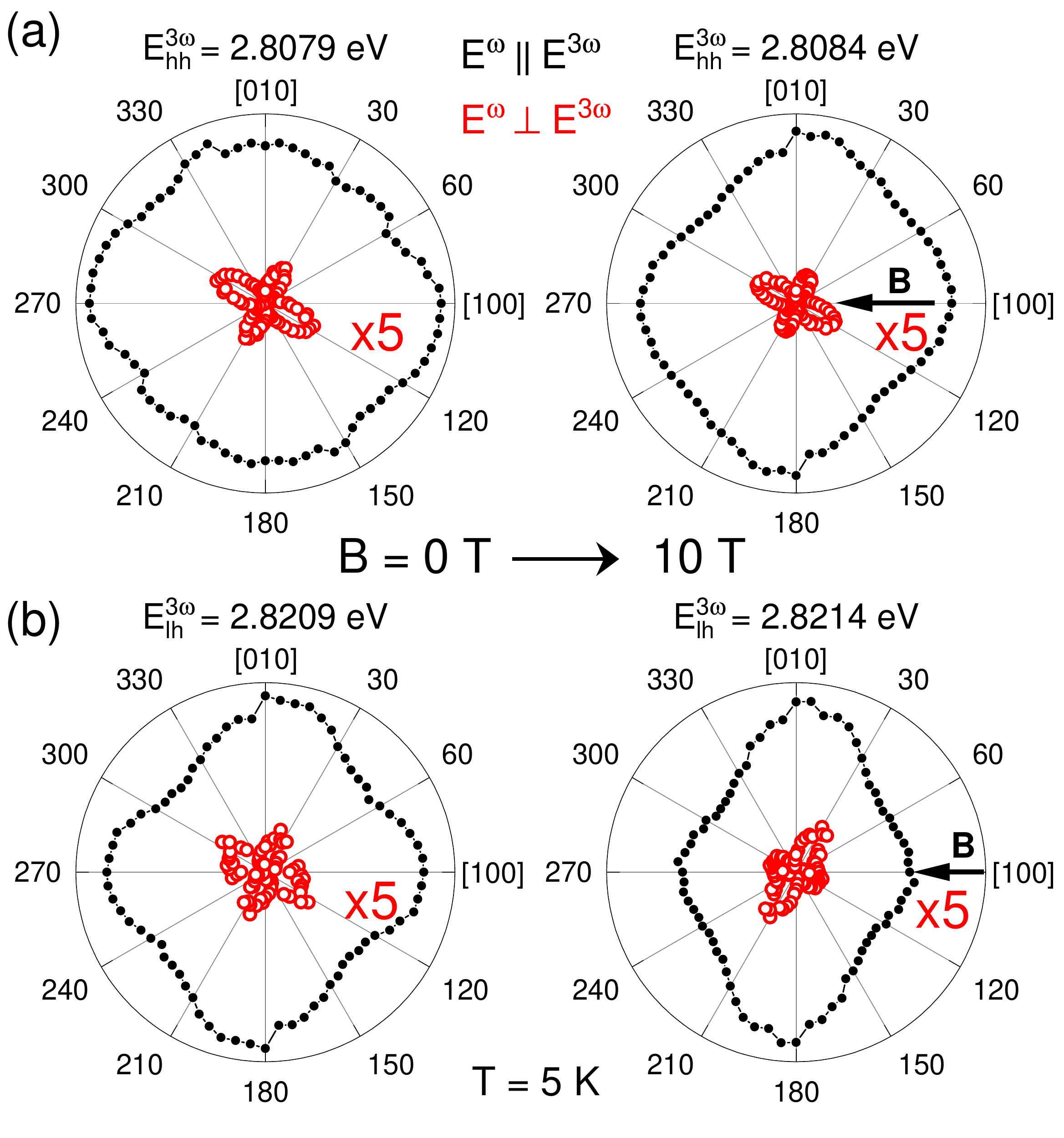}
		\caption{THG rotational anisotropies at $B=0$~T (left) and 10~T (right) of the (a) e1-hh1 exciton and (b) e1-lh1 exciton, as labeled in Fig.~\ref{pic.[001]_THG_all_ps+fs}. The solid (black) and open (red) circles show data for the $\textbf{E}^\omega\parallel\textbf{E}^{3\omega}$ and $\textbf{E}^\omega\perp\textbf{E}^{3\omega}$ (magnified by factor 5) polarization configurations, respectively.}
		\label{pic.[001]_THG_1S+b_0+10Tv_Ani}
	\end{center}
\end{figure}

THG rotational anisotropies of the 1S excitons involving the heavy-hole (e1-hh1) and light-hole (e1-lh1) transitions are shown in Figs.~\ref{pic.[001]_THG_1S+b_0+10Tv_Ani}(a) and \ref{pic.[001]_THG_1S+b_0+10Tv_Ani}(b), respectively. At zero magnetic field, the 1S(hh) anisotropy measured for parallel polarizers ($\textbf{E}^\omega\parallel\textbf{E}^{3\omega}$, the black points) has an almost isotropic shape, whereas the crossed anisotropy ($\textbf{E}^\omega\perp\textbf{E}^{3\omega}$, red circles) has much lower intensity and a shape with four lobes. In the magnetic field of 10~T, the parallel anisotropy is modified. It becomes elongated perpendicular to the magnetic field direction, i.e., along the [010] crystal axis. The perpendicular anisotropy does not show visible modifications. The parallel 1S(lh) anisotropy is not isotropic at zero magnetic field, but has a fourfold-symmetry shape with weak contrast modulation. In magnetic fields, it becomes elongated along the [010] crystal axis as well. The perpendicular anisotropy of 1S(lh) has a fourfold symmetry shape, the same as the 1S(hh) resonance, and is not modified by the magnetic field. 

Note that we observed similar shapes for the THG rotational diagrams for the 1S exciton in bulk ZnSe \cite{Mund20}. However, the theoretical description differs for bulk and MQW samples due to their different symmetries, as will be shown in Sec.~\ref{sec.discussion}.

\subsection{Second harmonic generation}
\label{sec.SHG}

\begin{figure}[h]
\begin{center}
	\includegraphics[width=0.5\textwidth]{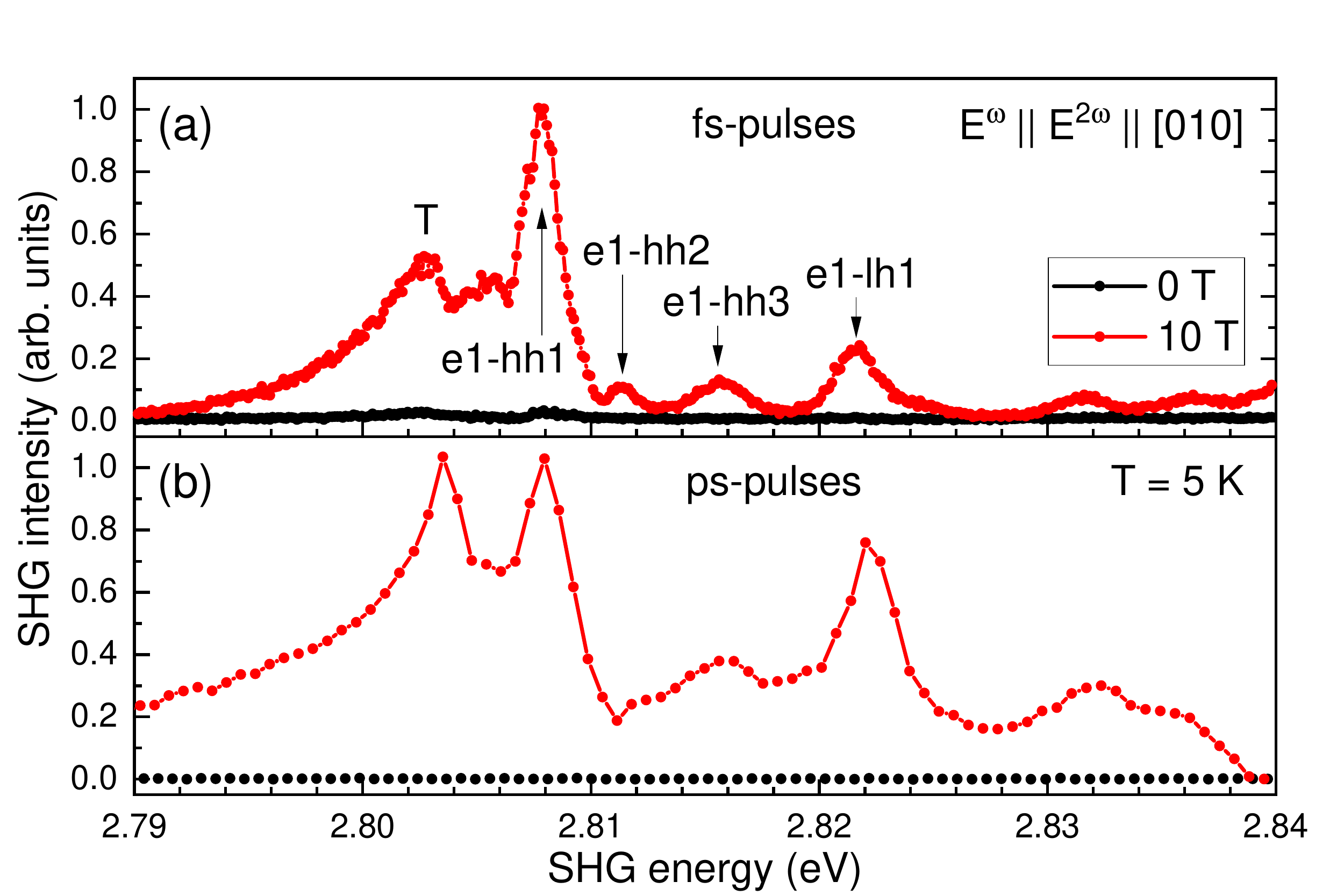}
	\caption{SHG spectra of ZnSe/BeTe MQW measured at $B=0$~T (black) and 10~T (red) for $T=5$~K in normal light incidence ($\theta=0^\circ$, $\textbf{k}^\omega\parallel [001]$, $\textbf{B}\perp\textbf{k}^\omega$). (a) Using femtosecond pulses centered at $2\hbar\omega=2.812$~eV. (b) With scanned picosecond pulses.}
	\label{pic.[001]_SHG_all_0+10Tv_fs+ps}
\end{center}
\end{figure}

Contrary to THG, SHG is symmetry forbidden in a ZnSe/BeTe MQW for light propagating along the [001] crystal direction. Indeed, we find no SHG signal in the range of exciton resonances at zero magnetic field, see the black symbols in Fig.~\ref{pic.[001]_SHG_all_0+10Tv_fs+ps}. However, by applying a magnetic field in the Voigt geometry ($\textbf{B}\perp\textbf{k}^\omega$), a manifold of resonances appears in the SHG spectrum (red dots). Their spectral positions coincide well with the exciton resonances in the reflectivity and THG spectra, compare with Figs.~\ref{pic.Refl+PL}(a) and \ref{pic.[001]_THG_all_ps+fs}. Similar to THG, the SHG spectra can be measured with fs lasers pulses or scanned ps pulses. The fs-pulse method provides better spectral resolution, compare the panels (a) and (b) of Fig.~\ref{pic.[001]_SHG_all_0+10Tv_fs+ps}. 

\FloatBarrier

\begin{figure}[h]
	\begin{center}
		\includegraphics[width=0.48\textwidth]{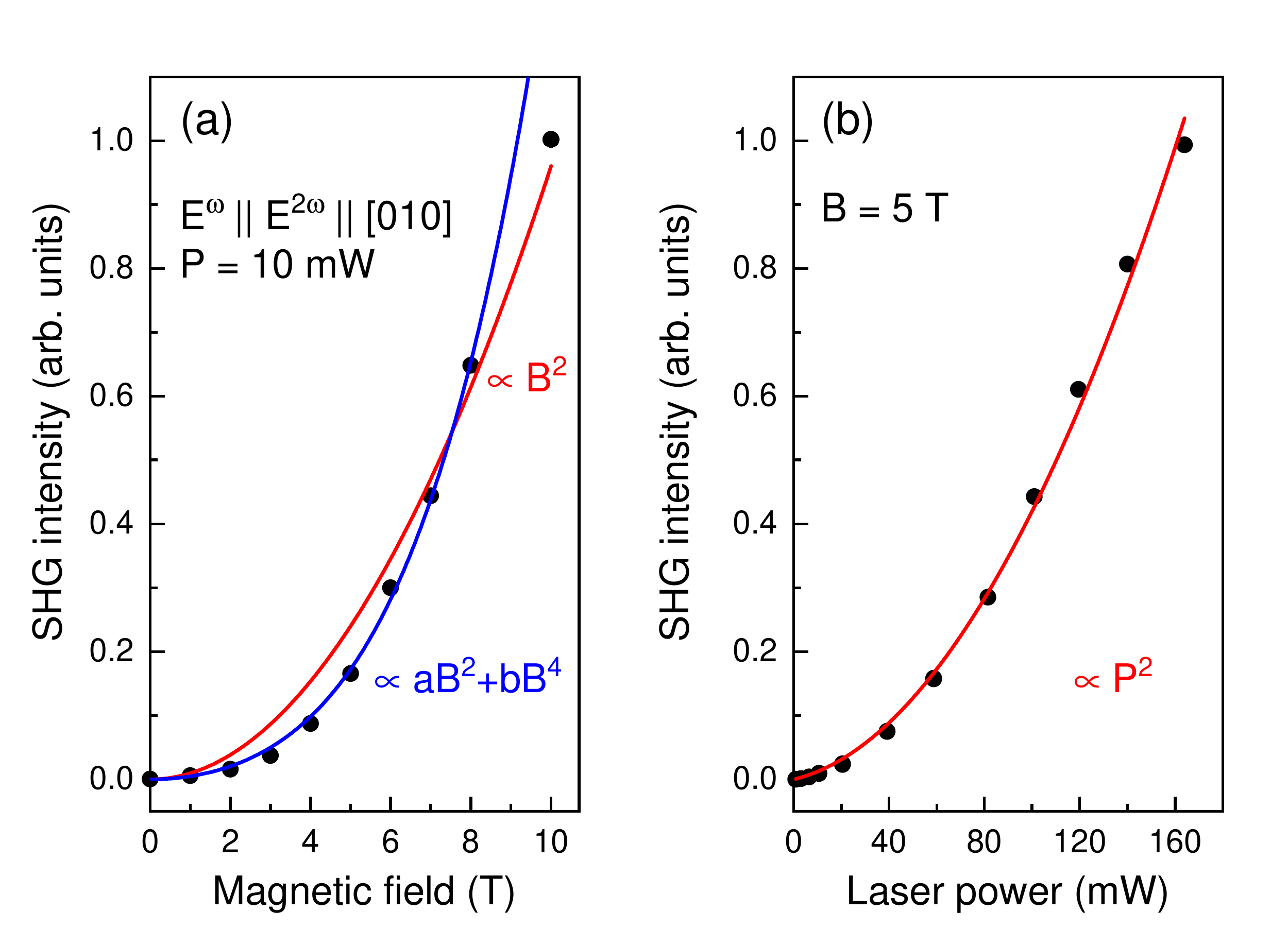}
		\caption{SHG intensity dependence of the e1-hh1 exciton resonance (a) on magnetic field for $P=10$~mW, and (b) on laser power at $B=5$~T. Symbols give the data and red lines show quadratic fits to the data. The blue line in panel (a) is a fit accounting $B^2$ and $B^4$ terms with respective parameters $a=42$ and $b=0.75$. In this case data point at $B=10$~T was excluded from the fitting data.}
		\label{pic.[001]_SHG_1S_0-10Tv_Int-B-P-dep}
	\end{center}
\end{figure}

In Figure~\ref{pic.[001]_SHG_1S_0-10Tv_Int-B-P-dep}(a), the SHG intensity of the e1-hh1 resonance is shown as a function of magnetic field. The red line is a fit with $I^{\mathrm{SHG}} = c B^2$. It has good agreement with the experimental data. For the magnetic-field-induced SHG signals on the exciton resonances that were reported for bulk semiconductors like GaAs, CdTe and ZnO~\cite{Yakovlev18,Lafrentz13,Sanger06}, a $B^2$ dependence is rather typical. In order to make this fit even better, we account for higher terms using $I^{\mathrm{SHG}} = a B^2 + b B^4$. One can see from the blue line in Fig.~\ref{pic.[001]_SHG_1S_0-10Tv_Int-B-P-dep}(a) that a very good agreement is achieved when the data point at $B=10$~T is not taken into account. The fit amplitudes of the respective terms are $a=42$ and $b=0.75$, showing that the $B^2$ term is of main importance. The presence of the $B^4$ term can be related to, e.g., the magnetic-field-induced mixing of heavy-hole and light-hole states.  

We have measured also the power dependence of the SHG intensity at $B=5$~T, which follows the $P^2$ dependence expected for a nonlinear process with two-photon excitation, see Fig.~\ref{pic.[001]_SHG_1S_0-10Tv_Int-B-P-dep}(b).    

\begin{figure}[h]
	\begin{center}
		\includegraphics[width=0.5\textwidth]{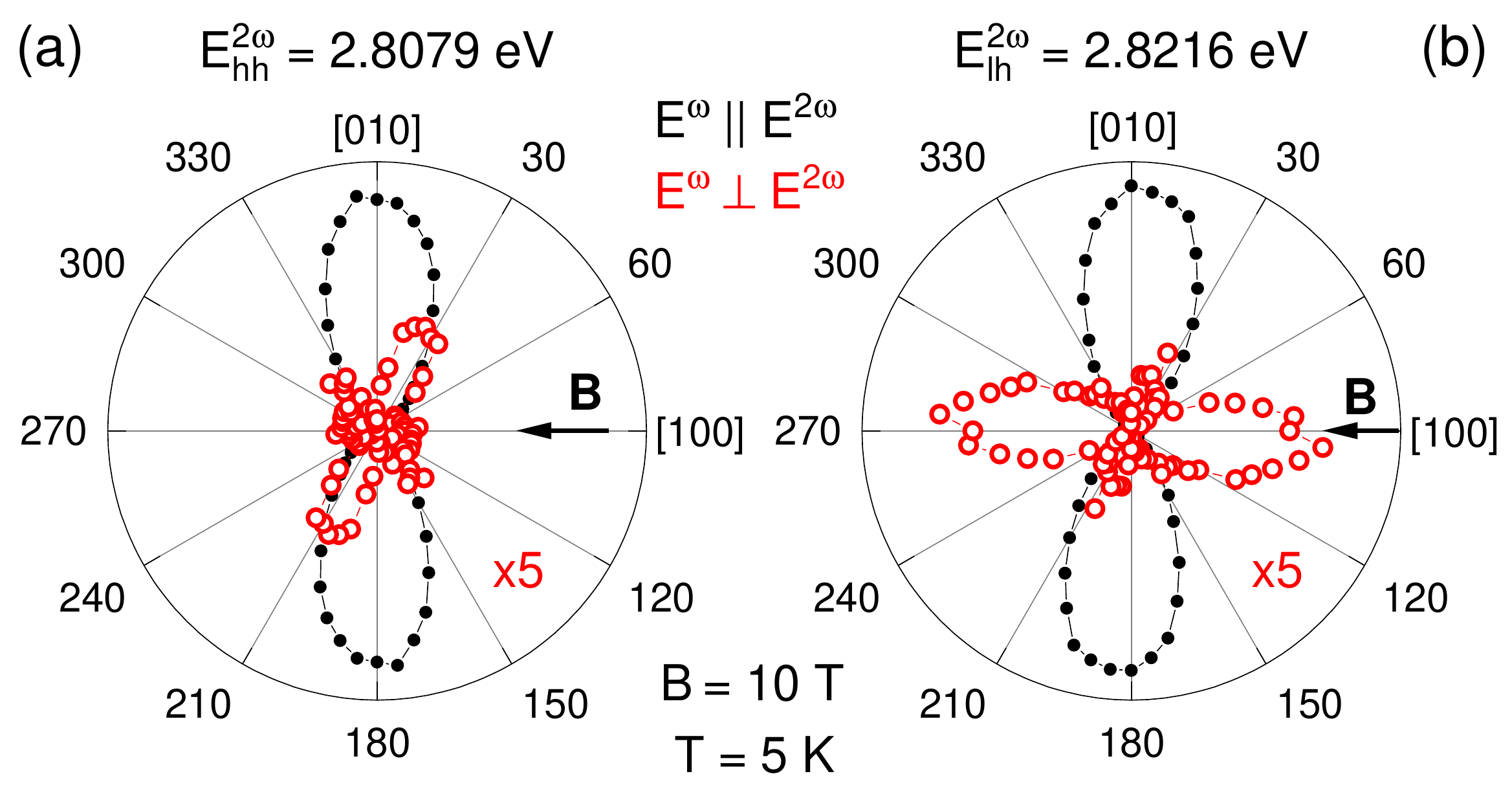}
		\caption{SHG rotational anisotropies recorded at the resonance energies of the e1-hh1 and e1-lh1 excitons, measured in a magnetic field of 10~T. Solid (black) and open (red) circles show data for the $\textbf{E}^\omega\parallel\textbf{E}^{2\omega}$ and $\textbf{E}^\omega\perp\textbf{E}^{2\omega}$ configurations, respectively.}
		\label{pic.[001]_SHG_1S+b_10Tv_Anis}
	\end{center}
\end{figure}

Rotational anisotropies of the e1-hh1 and e1-lh1 SHG lines are shown in Fig.~\ref{pic.[001]_SHG_1S+b_10Tv_Anis}. The parallel anisotropies ($\textbf{E}^\omega\parallel\textbf{E}^{2\omega}$) of both resonances have a twofold symmetry pattern with the lobes aligned along the [010] crystal direction, i.e. perpendicular to the magnetic field. The anisotropies measured with crossed polarizers ($\textbf{E}^\omega\perp\textbf{E}^{2\omega}$) have much weaker intensity (reduced by a factor of about $5-10$) and have different shapes. They show fourfold symmetries, but with different intensities of the lobes. The stronger component for the heavy-hole exciton is aligned along the direction $25^{\circ}$, while for the light-hole exciton is pointed along $90^{\circ}$.

\subsection{SHG from GaAs substrate in tilted geometry}
\label{sec.SHG_tilted}

\begin{figure}[h]
	\begin{center}
		\includegraphics[width=0.5\textwidth]{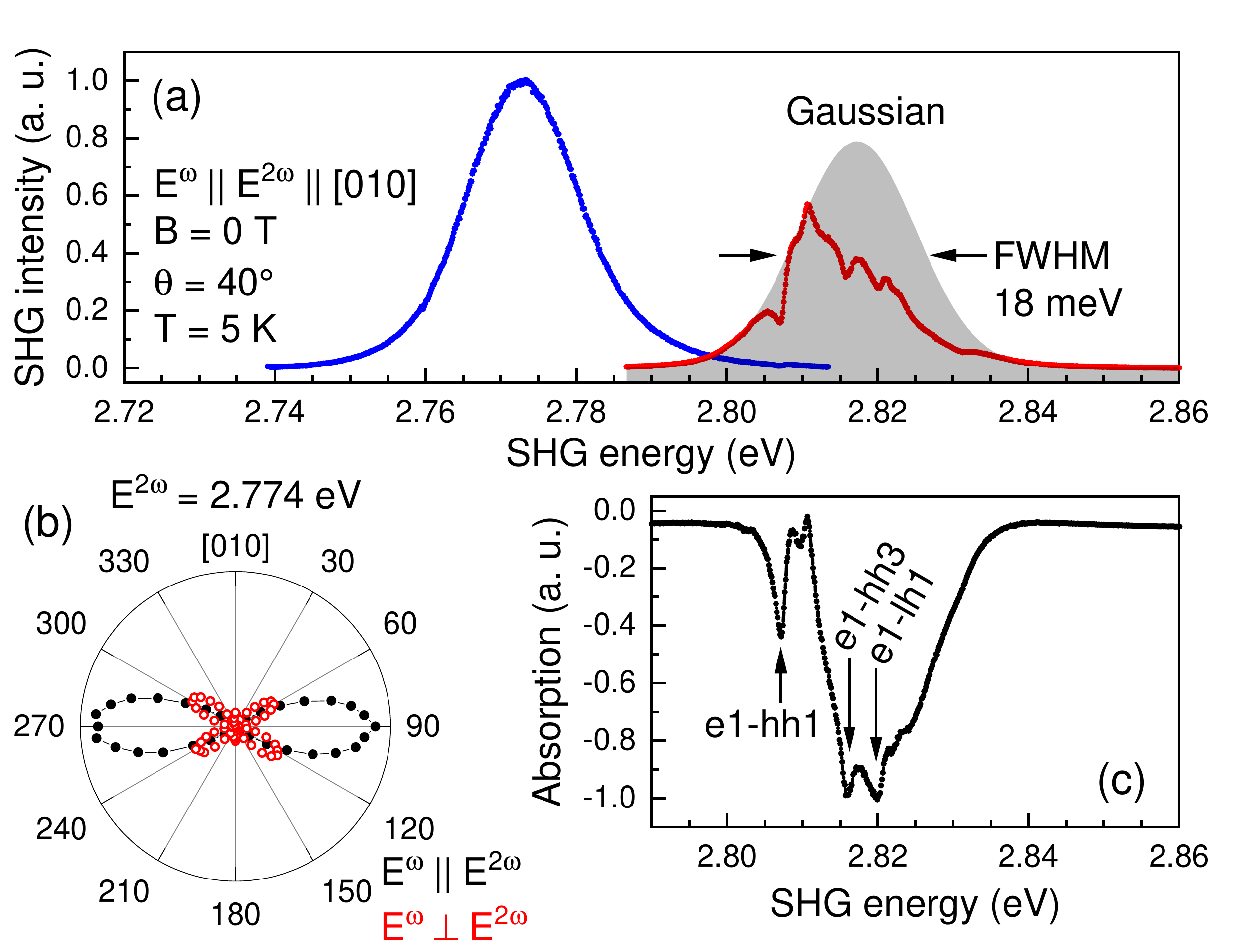}
		\caption{(a) SHG signal measured at tilted geometry ($\theta \approx 40^\circ$) using nonresonant (blue) and resonant (red) two-photon excitation. The gray shaded area represents a Gaussian fitted as an envelope to the red SHG signal. (b) Representative rotational anisotropy of SHG signal measured below the QWs exciton resonances at $2\hbar\omega=2.774$~eV. Solid (black) and open (red) circles are for the $\textbf{E}^\omega\parallel\textbf{E}^{2\omega}$ and $\textbf{E}^\omega\perp\textbf{E}^{2\omega}$ polarization configurations, respectively. (c) Difference intensity between the red SHG spectrum and the Gaussian fit shown in panel (a). It presents the absorption of the SHG generated in the GaAs substrate by the QW excitons.}
		\label{pic.[001]_SHG_all_tilt}
	\end{center}
\end{figure}

SHG signal becomes symmetry-allowed and can be observed at zero magnetic field when the incident laser light $\textbf{k}^\omega_2$ propagates along a direction different from the $[001]$ crystal direction, i.e. when the crystal is tilted~\cite{Sanger06}.  We set the tilting angle to $\theta\approx40^\circ$, corresponding to an internal angle between the light propagation direction $\textbf{k}^\omega$ and the $[001]$ crystal direction of $\approx13^\circ$. The corresponding SHG spectra are shown in Fig.~\ref{pic.[001]_SHG_all_tilt}(a). The blue line shows the spectrum measured with fs-pulses centered at $2\hbar\omega=2.774$~eV, which is below the exciton resonances in the MQW, but above the band gap of the GaAs substrate at 1.52~eV. The spectrum has the Gaussian shape of the femtosecond pulse spectrum. This signal corresponds to SHG generated in the GaAs substrate. When the laser energy is shifted to the MQW exciton resonances at $2\hbar\omega=2.816$~eV, the spectral shape becomes modulated, showing minima at the energies of the MQW exciton resonances. This is seen better in Fig.~\ref{pic.[001]_SHG_all_tilt}(c), where we subtract the Gaussian laser profile from the red spectrum in Fig.~\ref{pic.[001]_SHG_all_tilt}(a).   

We explain this finding by the generation of SHG light in the GaAs substrate and its subsequent absorption by excitons in the quantum wells. The substrate has cubic T$_d$ symmetry where SHG is forbidden for $\mathbf{k}^\omega_2\parallel[001]$ in absence of external fields, but SHG becomes allowed for a tilted sample. The band gap energy of GaAs is 1.52~eV at low temperatures. Therefore, two photons with $2\hbar\omega=2.816$~eV sum energy excite the continuum far above the GaAs band gap, leading to SHG there. The SHG light passes the quantum well layers, where it can be absorbed by the MQW excitons in a one photon process.

In principle, there could be a substrate contribution to the signal for SHG in magnetic field and in THG, too. We consider this contributions to have low intensity compared to the quantum well excitons. Furthermore, the SHG and THG from the substrate should show no distinct resonances at energies far above the band gap in the continuum states of GaAs. In Figs.~\ref{pic.[001]_THG_all_ps+fs} and \ref{pic.[001]_SHG_all_0+10Tv_fs+ps}, the signal intensity drops to close to zero outside the exciton resonance range. This proves that the substrate contributions can be neglected for measurements at normal light incidence ($\mathbf{k}^\omega_2\parallel[001]$).

\section{Discussion and modeling of rotational anisotropies}
\label{sec.discussion}

In this section, we analyze the SHG and THG rotational diagrams. We apply group theory and simulate the angle dependent signal intensity. This is done by considering the symmetries of the exciton states, the incoming and outgoing photons and the applied magnetic field.

\begin{figure}[h]
	\begin{center}
		\includegraphics[width=0.48\textwidth]{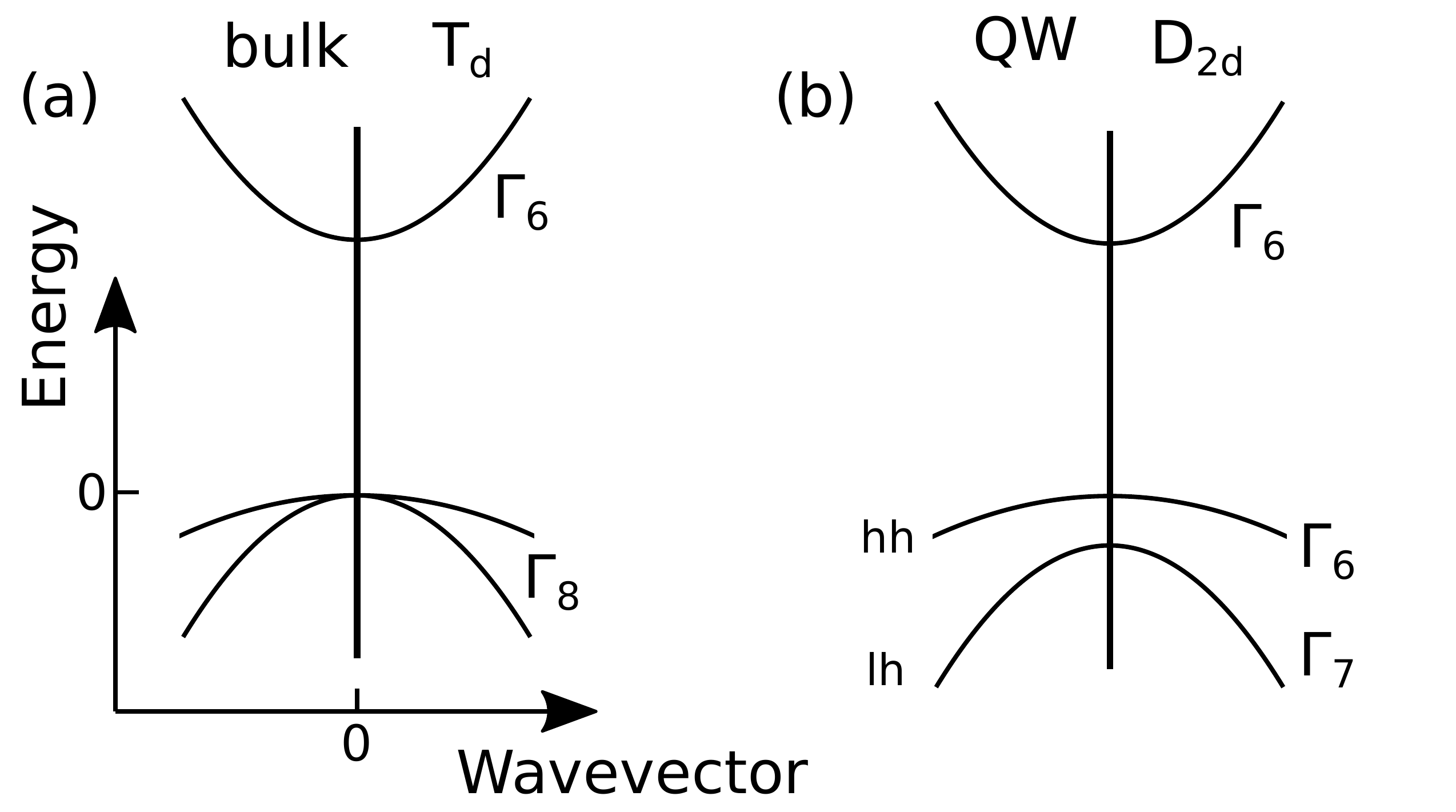}
		\caption{Irreducible representations of the conduction and valence bands: point groups of (a) bulk material, $T_{d}$, and (b) quantum wells, $D_{2d}$. The spin-orbit split valence band at lower energy is not shown.}
		\label{pic.irr-reps_D2d_C2v}
	\end{center}
\end{figure}

The exciton symmetries in the ZnSe layers are obtained from the symmetries of the associated conduction and valence bands and the exciton envelope, which is, however, simply $\Gamma_1^+$ for 1S excitons. Whereas ZnSe bulk material has the point symmetry group $T_d$, the symmetry of the quantum well structure is reduced to $D_{2d}$. Note, that in the studied ZnSe/BeTe MQW the two interfaces of each ZnSe quantum well are formed by Zn-Te bonds, which prevents a further possible symmetry reduction to $C_{2v}$, as could be the case for nonequivalent Zn-Te and Be-Se interfaces \cite{Yakovlev00}. The strain and quantum confinement induce a splitting of the heavy-hole ($\Gamma_6$) and light-hole ($\Gamma_7$) bands, which changes their irreducible representations, see Fig.~\ref{pic.irr-reps_D2d_C2v}.

\FloatBarrier

The symmetry of an exciton ($\Gamma_\textrm{exc}$) is determined by the product of the irreducible representations of the electron ($\Gamma_\textrm{e}$), the hole ($\Gamma_\textrm{h}$) and the envelope of the relative motion of both ($\Gamma_\textrm{env}$):
\begin{equation}
\Gamma_\textrm{exc}=\Gamma_\textrm{e}\otimes\Gamma_\textrm{h}\otimes\Gamma_\textrm{env}.
\end{equation}
In the case of $D_{2d}$ symmetry, the 1S exciton states involving either a heavy-hole ($\Gamma_6$) or a light-hole ($\Gamma_7$) are given by
\begin{eqnarray}\label{eq.exc-states_hh}
\Gamma_\textrm{hh}&=&\Gamma_6\otimes\Gamma_6\otimes\Gamma_1=\Gamma_1\oplus\Gamma_2\oplus\Gamma_5,\\\label{eq.exc-states_lh}
\Gamma_\textrm{lh}&=&\Gamma_6\otimes\Gamma_7\otimes\Gamma_1=\Gamma_3\oplus\Gamma_4\oplus\Gamma_5.
\end{eqnarray}
The required tables for the calculation can be found in Koster $et~al.$ \cite{Koster}.

Next, we calculate which of the exciton states in Eqs.~(\ref{eq.exc-states_hh}) and (\ref{eq.exc-states_lh}) can be excited by one, two, or three photons. In the point group $D_{2d}$, a single photon that is polarized in the quantum well plane (i.e. in the $(xy)$ plane) is described by $\Gamma_5$ symmetry.

For SHG and THG, the addressed exciton states need to be allowed for excitation by two and three photons, respectively. Two photons can excite states of the symmetries
\begin{equation}
\Gamma_5\otimes\Gamma_5=\Gamma_1\oplus\Gamma_2(=0)\oplus\Gamma_3\oplus\Gamma_4.
\end{equation}
Note that $\Gamma_2$ can not be excited by two photons of the same polarization. Using three photons, only states of $\Gamma_5$ symmetry can be excited because every state symmetry that can be excited by two photons, multiplied by the symmetry of the third photon results in $\Gamma_5$ symmetry,
\begin{equation}
\Gamma_5\otimes\Gamma_5\otimes\Gamma_5=3\Gamma_5.
\end{equation}

In order to calculate rotational anisotropies, the amplitudes, $\textrm{E}^\omega_N(\psi)$, of the incoming photons are described by the polarizer angle $\psi$, with $N=2$ and 3 for SHG and THG. The amplitudes of the outgoing photons, $\textrm{E}^{N\omega}(\varphi)$, are dependent on the analyzer angle $\varphi$. For $\textbf{k}^\omega_N\parallel[001]$ and, respectively, $\textbf{k}^{N\omega}\parallel[001]$  the amplitude of the outgoing one-photon process  is given by
\begin{equation}
\textbf{E}^{N\omega}(\varphi)=
\begin{pmatrix}
-\textrm{sin}(\varphi)\\\phantom{-}\textrm{cos}(\varphi) \\ 0
\end{pmatrix}.
\end{equation}
For consistency with previous publications and the multi-photon operators given below, we define the one-photon emission operator as $O_\textrm{out}(\textrm{E}^{N\omega},\varphi)=\textbf{E}^{N\omega}(\varphi)$.

The two- and three-photon excitation operators are given by
\begin{eqnarray}
O_\textrm{in}(\textbf{E}^\omega_2,\psi)&=&\textbf{E}^\omega_2(\psi)\otimes\textbf{E}^\omega_2(\psi),\\
O_\textrm{in}(\textbf{E}^\omega_3,\psi)&=&\textbf{E}^\omega_3(\psi)\otimes\textbf{E}^\omega_3(\psi)\otimes\textbf{E}^\omega_3(\psi).
\end{eqnarray}
The total, angle dependent SHG and THG intensities are calculated by
\begin{eqnarray}
I^\textrm{SHG}&\propto&|O_\textrm{in}(\textbf{E}^\omega_2,\psi)O_\textrm{out}(\textbf{E}^{2\omega},\varphi)|^2,\\
I^\textrm{THG}&\propto&|O_\textrm{in}(\textbf{E}^\omega_3,\psi)O_\textrm{out}(\textbf{E}^{3\omega},\varphi)|^2.
\end{eqnarray}
The rotational anisotropies for parallel ($\textbf{E}^\omega\parallel\textbf{E}^{N\omega}$) and crossed ($\textbf{E}^\omega\perp\textbf{E}^{N\omega}$) polarization configuration are obtained by setting either $\varphi=\psi$ or $\varphi=\psi+90^\circ$.

\subsection{SHG rotational anisotropies}

From the fact that two photons can excite only states of $\Gamma_1$, $\Gamma_3$, or $\Gamma_4$ symmetry, whereas one photon can only be emitted by states of $\Gamma_5$ symmetry, we can already conclude that no SHG signal is expected for $\textbf{k}^\omega_2\parallel[001]$.

\begin{figure}[h]
	\begin{center}
		\includegraphics[width=0.48\textwidth]{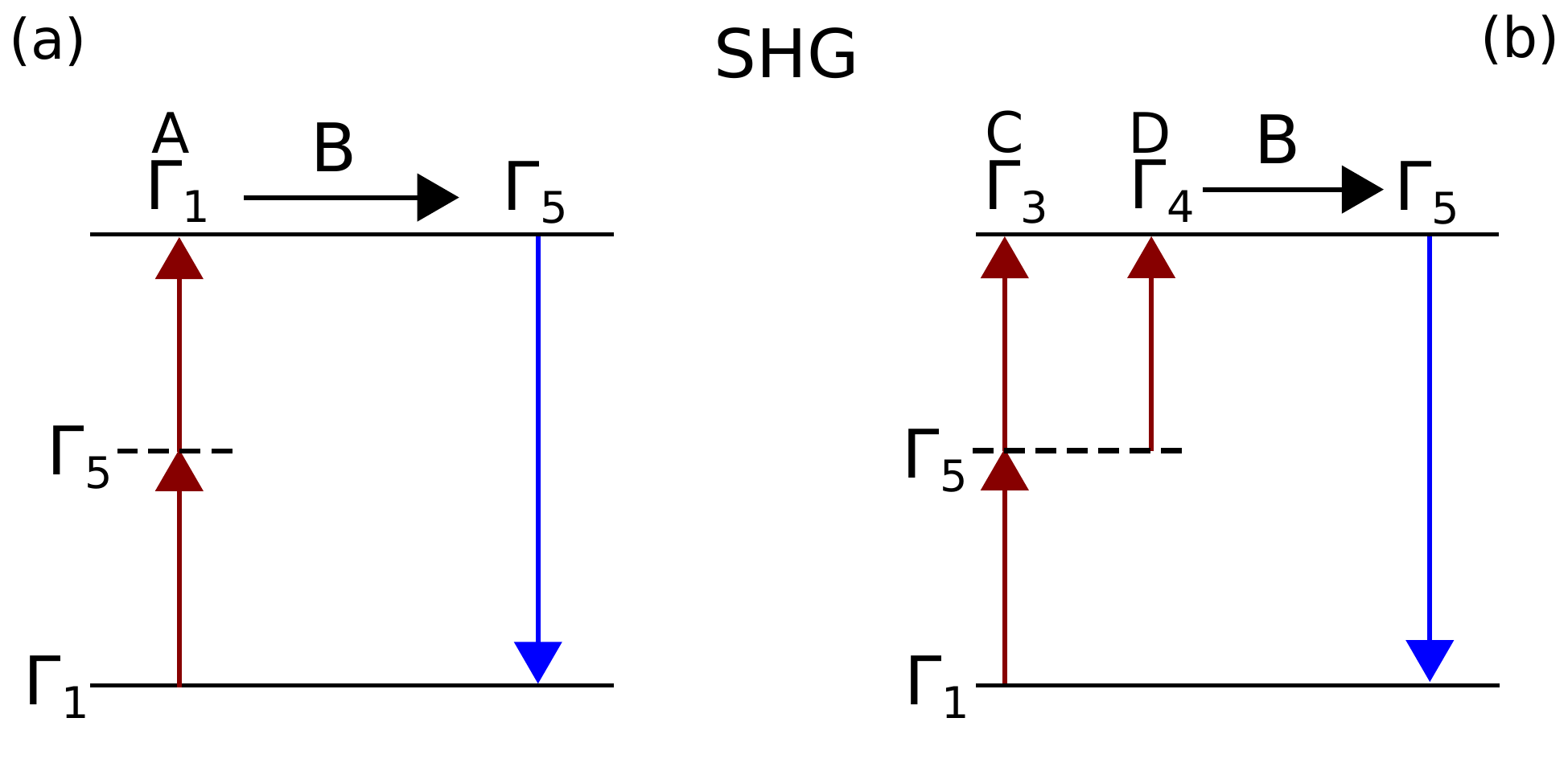}
		\caption{Two-photon excitation (red arrows) and one-photon emission (blue arrows) paths of 1S excitons in SHG. The dashed lines represent intermediate states with corresponding symmetry. The red and blue arrows are photons of $\Gamma_5$ symmetry. The black horizontal arrow represents mixing of states by a magnetic field. SHG processes for (a) the heavy-hole 1S exciton in path 'A' and (b) the light-hole excitons through paths 'C' and 'D'.}
		\label{pic.SHG+B_processes}
	\end{center}
\end{figure}

The situation changes when a magnetic field is applied along the $x$-axis ([100] crystal direction, Voigt geometry, $\textbf{B}\perp\textbf{k}^\omega$). The magnetic field has $\Gamma_5$ symmetry and couples (i) the heavy-hole states of symmetry $\Gamma_1$ to $\Gamma_5$ and (ii) the light-hole states of symmetries $\Gamma_3$ and $\Gamma_4$ to $\Gamma_5$, see Eqs.~\eqref{eq.exc-states_hh} and \eqref{eq.exc-states_lh}. Therefore, (i) two photons excite a $\Gamma_1$-state ($O_\textrm{in1}$) which is coupled to $\Gamma_5$ and can emit a photon. (ii) Two photons excite $\Gamma_3$ or $\Gamma_4$ states ($O_\textrm{in3, 4}$) that are both coupled to $\Gamma_5$ which can emit again one photon. Both processes are visualized in Fig.~\ref{pic.SHG+B_processes}.

\begin{figure}[h]
	\begin{center}
		\includegraphics[width=0.48\textwidth]{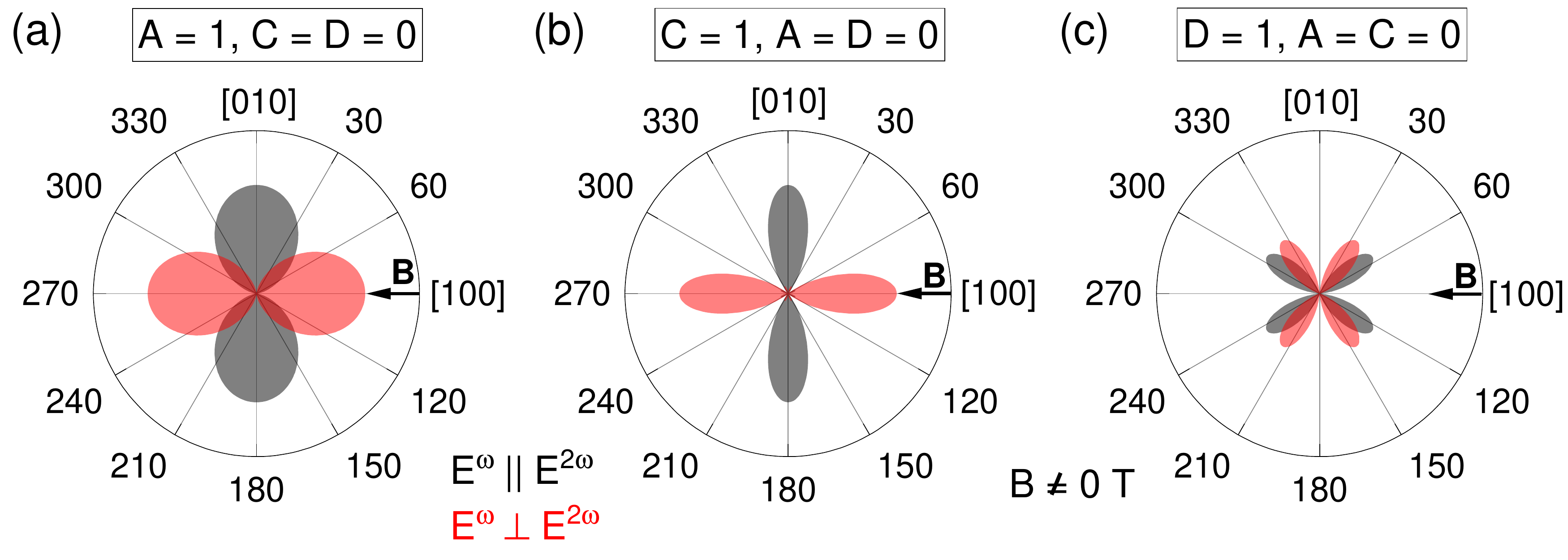}
		\caption{SHG rotational anisotropies in magnetic field for the individual SHG paths $A$, $C$ and $D$, respectively in panels (a), (b) and (c). The direction of the magnetic field is indicated by the small black arrow with the boldface letter $\textbf{B}$ on top. The gray and red shaded areas show simulations for $\textbf{E}^\omega\parallel\textbf{E}^{2\omega}$ and $\textbf{E}^\omega\perp\textbf{E}^{2\omega}$ configurations, respectively.}
		\label{pic.SHG+B_Ani_Simu_Comps}
	\end{center}
\end{figure}

Although the heavy-hole and light-hole 1S states are split by several meV we find in our simulation that interference of the paths $A$, $C$ and $D$ is necessary to model the rotational anisotropies. Therefore, it is necessary to add up the angle-dependent amplitudes of each path prior to calculating the square of this sum, in order to obtain the SHG intensity. The magnetic field induced SHG is given by
\begin{equation}
I^\textrm{SHG}_\textrm{B}\propto B^2|(AO_\textrm{in1}+CO_\textrm{in3}+DO_\textrm{in4})O_\textrm{out}|^2
\end{equation}
with the parameters $A$, $C$ and $D$ are for the individual paths. The explicit expressions for parallel and crossed anisotropies are
\begin{eqnarray}\label{eq.SHG+B1}
I^\textrm{SHG}_{\textrm{B, }\parallel}&\propto& B^2\textrm{sin}(\psi)^2\left[A+D+(C+D)\textrm{cos}(2\psi)\right]^2,\\ \label{eq.SHG+B2}
I^\textrm{SHG}_{\textrm{B, }\perp}&\propto& B^2\textrm{cos}(\psi)^2\left[A-D+(C+D)\textrm{cos}(2\psi)\right]^2.
\end{eqnarray}
To show the importance of this interference between the three paths, we give in Fig.~\ref{pic.SHG+B_Ani_Simu_Comps} the results for each single path. Thus, in panels (a)-(c) one of the three parameters is set to unity, while the other two are set to zero.

\begin{figure}[h]
	\begin{center}
		\includegraphics[width=0.48\textwidth]{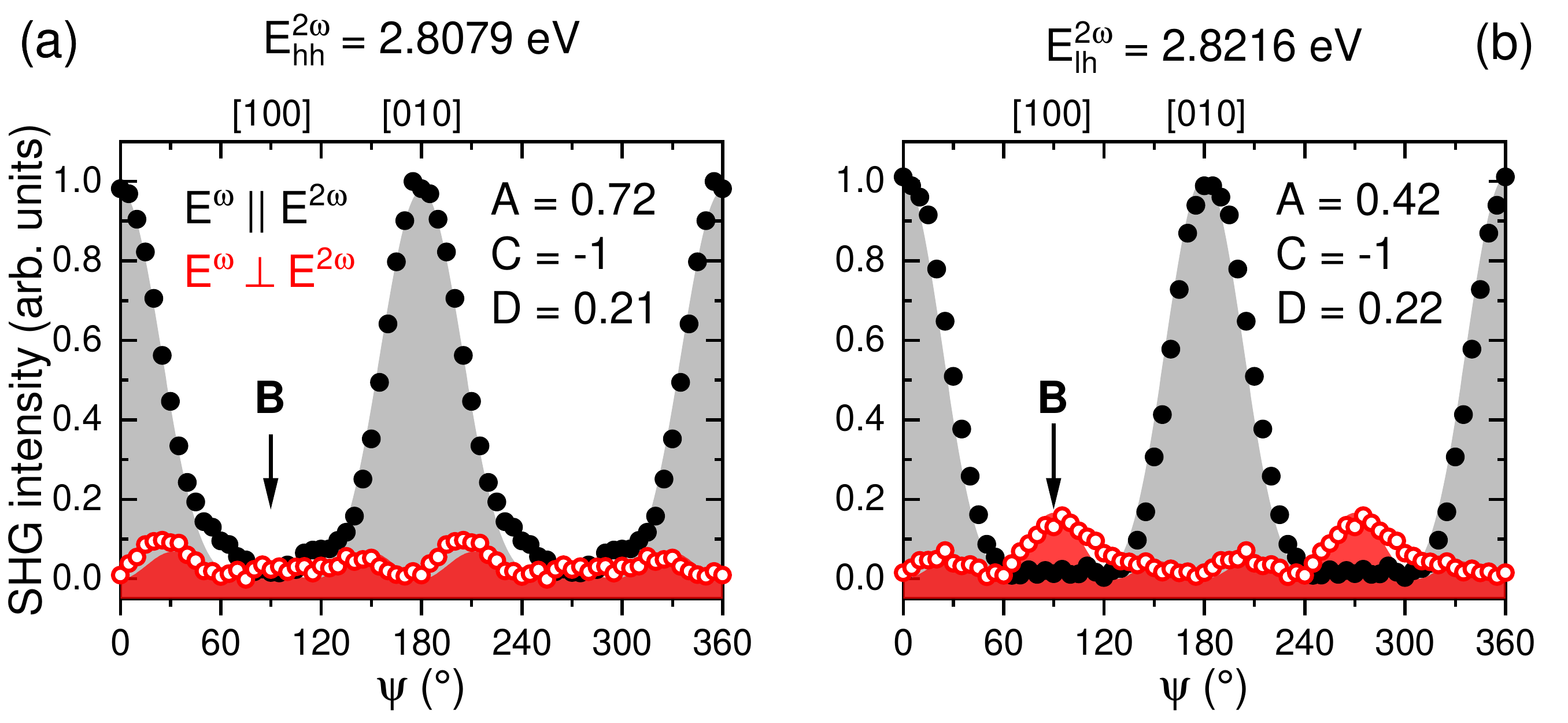}
		\caption{Experimental SHG data and simulations for the (a) heavy-hole and (b) light-hole 1S exciton at $B=10$~T. The small black arrow with the boldface letter $\textbf{B}$ on top indicate the direction of the magnetic field. Solid (black) and open (red) circles represent data for the $\textbf{E}^\omega\parallel\textbf{E}^{2\omega}$ and $\textbf{E}^\omega\perp\textbf{E}^{2\omega}$ polarization configurations, respectively. The gray and red shaded areas show the corresponding simulations.}
		\label{pic.SHG+B_Ani_Simu}
	\end{center}
\end{figure}

Equations~(\ref{eq.SHG+B1}) and (\ref{eq.SHG+B2}) are fitted to the heavy-hole and light-hole 1S SHG anisotropies in Fig.~\ref{pic.[001]_SHG_1S+b_10Tv_Anis}. The comparison of the model calcualtions with the experimental data as well as the fit parameters are given in Fig.~\ref{pic.SHG+B_Ani_Simu}. We use here cartesian instead of polar coordinates for better visibility.

\subsection{THG rotational anisotropies}

\begin{figure}[h]
	\begin{center}
		\includegraphics[width=0.48\textwidth]{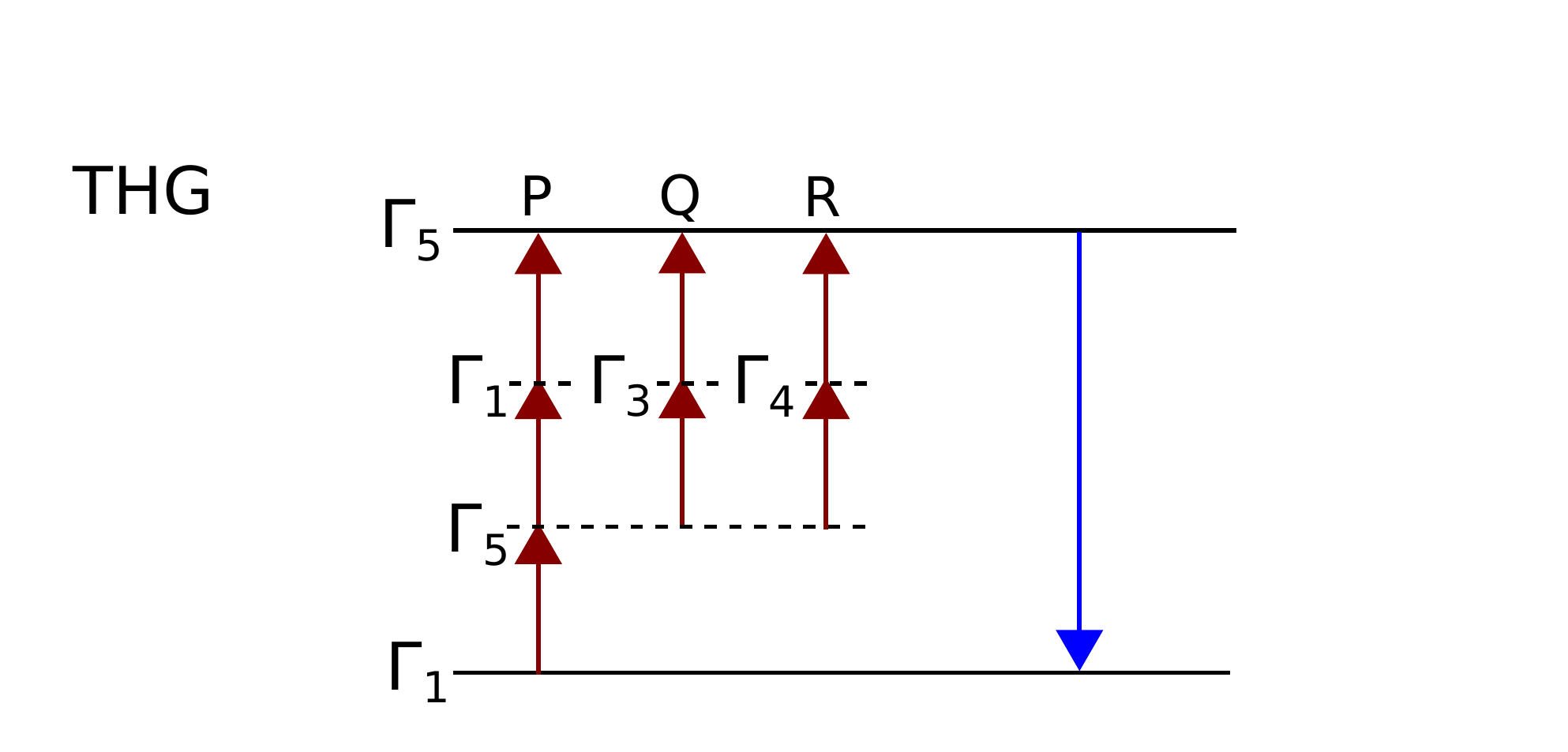}
		\caption{Three-photon excitation (red arrows) and one-photon emission (blue arrows) paths of the 1S $\Gamma_5$ exciton states in THG. The dashed lines represent intermediate states with corresponding symmetries. The red and blue arrows are photons of $\Gamma_5$ symmetry.}
		\label{pic.THG_processes}
	\end{center}
\end{figure}

In THG, signal is already allowed without application of an external magnetic field. The possible excitation and emission paths are presented in Fig~\ref{pic.THG_processes}.

The three paths add up to the total excitation of the $\Gamma_5$ 1S state. Thus, the THG signal is given by
\begin{equation}
I^\textrm{THG}\propto|(PO_\textrm{in1}+QO_\textrm{in3}+RO_\textrm{in4})O_\textrm{out}|^2
\end{equation}
with the parameters $P$, $Q$ and $R$ for describing the contributions of the individual paths.

\begin{figure}[h]
	\begin{center}
		\includegraphics[width=0.48\textwidth]{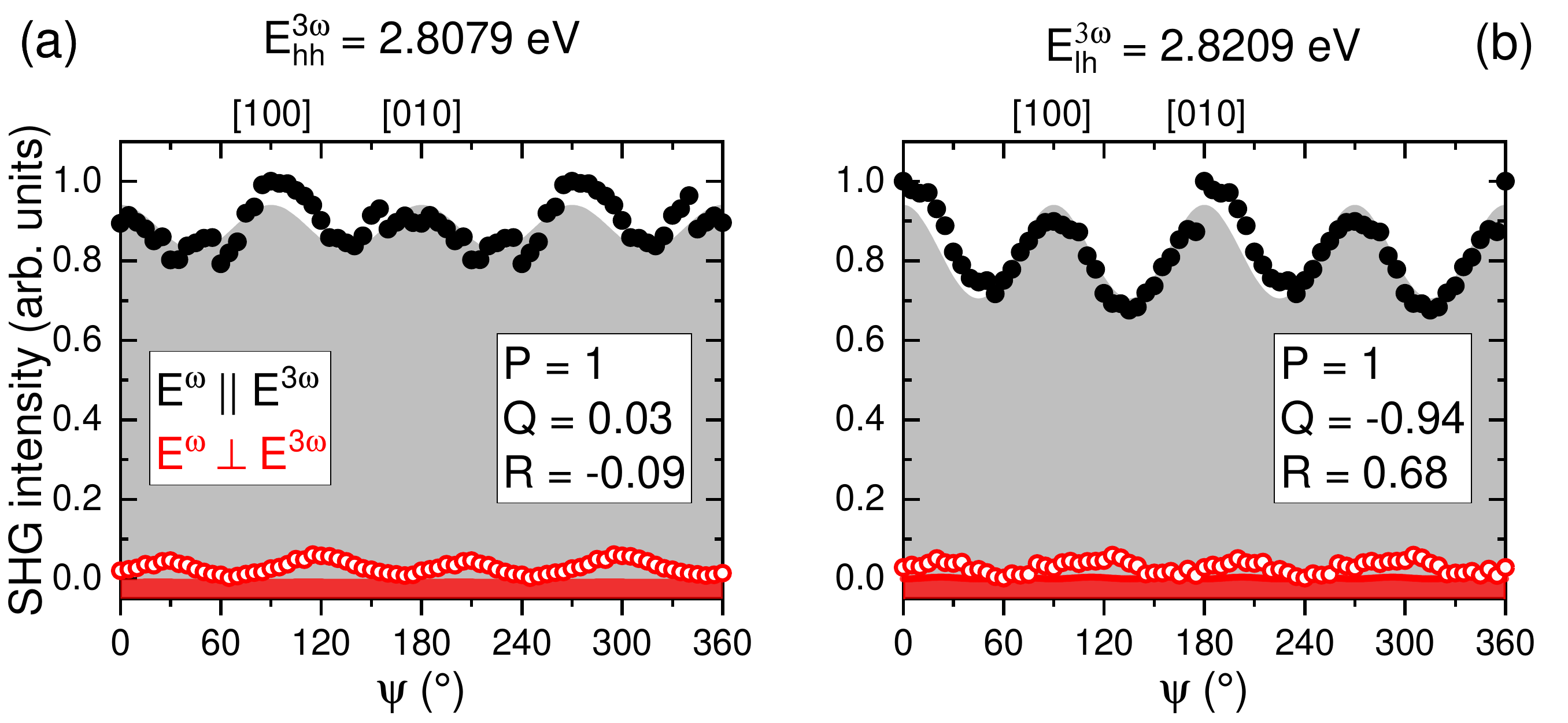}
		\caption{Experimental THG data and simulations for the (a) heavy-hole and (b) light-hole 1S exciton at $B=0$~T. The solid (black) and open (red) circles represent data for the $\textbf{E}^\omega\parallel\textbf{E}^{3\omega}$ and $\textbf{E}^\omega\perp\textbf{E}^{3\omega}$ polarization configurations, respectively. The gray and red shaded areas show corresponding simulations.}
		\label{pic.THG_Ani_Simu}
	\end{center}
\end{figure}

The expressions for the parallel and crossed anisotropies are
\begin{eqnarray}\label{eq.THG_par}
I^\textrm{THG}_\parallel&\propto&\frac{1}{16}\left[-2P+Q-R+(Q+R)\textrm{cos}(4\psi)\right]^2,\\\label{eq.THG_perp}
I^\textrm{THG}_\perp&\propto&\frac{1}{16}(Q+R)^2\textrm{sin}(4\psi)^2.
\end{eqnarray}
Equations~(\ref{eq.THG_par}) and (\ref{eq.THG_perp}) are fitted to the heavy-hole and light-hole 1S THG anisotropies at $B=0$~T in Fig.~\ref{pic.[001]_THG_1S+b_0+10Tv_Ani}(a). The results are shown in Fig.~\ref{pic.THG_Ani_Simu} in combination with the fit parameter values.

\FloatBarrier

On one hand, the small modulation of the almost constant heavy-hole 1S signal with a fourfold symmetry pattern in the parallel configuration can be reproduced by the simulation. Further, the stronger modulation for the light-hole 1S state is expressed by larger values of the fit parameters $Q$ and $R$. On the other hand, the low intensity signal with fourfold symmetry in the crossed configuration is not reproduced by the fit. This deviation might be explained by small sample misalignments and/or internal strain in the quantum wells. 

The application of a magnetic field will, unlike in the SHG case, not allow for the activation of additional excitation paths in THG. Only the $\Gamma_5$ states can be excited by three photons. Therefore, a possible mixing of exciton states by the magnetic field will not result in new contributions. However, the oscillator strength of the $\Gamma_5$ state might be transferred to dark states, reducing the emission intensity from $\Gamma_5$ components.

\begin{figure}[h]
	\begin{center}
		\includegraphics[width=0.48\textwidth]{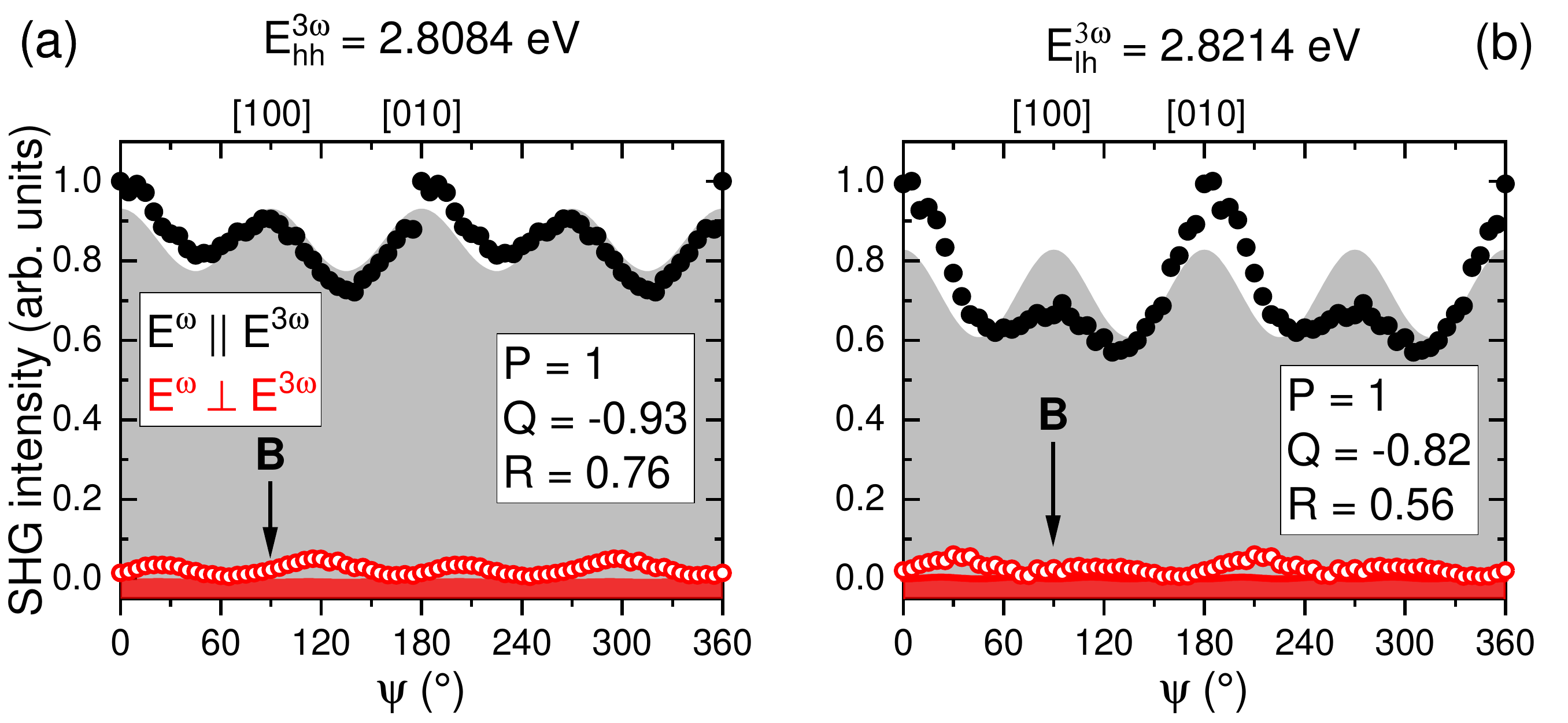}
		\caption{Experimental THG data and simulation for the (a) heavy-hole and (b) light-hole 1S exciton in the magnetic field of 10~T. The short black arrow with the boldface letter $\textbf{B}$ on top indicate the magnetic field direction. The solid (black) and open (red) circles represent data for the $\textbf{E}^\omega\parallel\textbf{E}^{3\omega}$ and $\textbf{E}^\omega\perp\textbf{E}^{3\omega}$ polarization configurations, respectively. The gray and red shaded areas show corresponding simulations.}
		\label{pic.THG+B_Ani_Simu}
	\end{center}
\end{figure}

Equations~(\ref{eq.THG_par}) and (\ref{eq.THG_perp}) are fitted to the heavy-hole and light-hole 1S THG anisotropies at $B=10$~T in Fig.~\ref{pic.[001]_THG_1S+b_0+10Tv_Ani}(b). The results are shown in Fig.~\ref{pic.THG+B_Ani_Simu} together with the used fit parameters.

\FloatBarrier

The parallel anisotropy of the heavy-hole 1S exciton can be well described whereas, the corresponding light-hole signal with the different intensities of the fourfold symmetry modulation cannot be matched. We assign this finding to one of several possible effects of the magnetic field. One can consider on that respect to a coupling of the [100] exciton component to a dark exciton state and, therefore, to a reduction of oscillator strength. Vice versa, the oscillator strength could also be increased by state couplings for certain polarizations, e.g., along [010] in Fig.~\ref{pic.THG+B_Ani_Simu}(b). Two-photon allowed excited heavy-hole states, energetically close to the light-hole 1S state, could become admixed and increase the THG intensity. The different intensities of the [100] and [010] exciton components could also originate from the quantum well interfaces. The two interfaces of the ZnSe quantum wells are terminated by Zn-Te bonds which are oriented perpendicular to each other \cite{Yakovlev00}, resulting in the different behaviour of the $x$- and $y$-components.

\section{Conclusions}

We have measured optical SHG and THG spectra of exciton resonances in a type-II ZnSe/BeTe quantum well structure. SHG signal for the light vector $\textbf{k}^\omega$ parallel to the growth direction of the sample ([001] crystal axis) is only allowed if an external magnetic field is applied. THG, on the other hand, is allowed in absence of the magnetic field. The comparison of excitation by fs- and ps-pulses shows that measurements using the spectrally broad fs-pulses provide a fast method to observe exciton resonances with high resolution. A symmetry analysis using group theory shows that the interference of several excitation paths has to be considered to simulate the measured rotational anisotropy diagrams in SHG and THG. A difference in the fitting parameters of THG rotational anisotropies has been found for the heavy- and light-hole 1S excitons, which could allow one to assign further higher lying resonances dominantly to the heavy-hole or light-hole band. The change of the THG anisotropy of the light-hole exciton with increasing magnetic field indicates higher order effects of the magnetic field on the exciton states, which may be related to band mixing.

\section{Acknowledgments}
The authors are thankful to E. L. Ivchenko, M. A. Semina, M. M. Glazov, and D. Fr\"ohlich for stimulating discussions. We acknowledge the financial support by the Deutsche Forschungsgemeinschaft through the Collaborative Research Centres TRR142 (Project B01) and TRR160 (project C8). Sample for this study has been grown by AW being working in W\"urzburg University, Germany.

\end{document}